# Electrically-tunable graphene nanomechanical resonators

Yi-Bo Wang[1,2,3]†, Zhuo-Zhi Zhang[1,2,3]†, Chen-Xu Wu[1,2,3], Yu-Shi Zhang[1,2,3], Guo-Sheng Lei[1,2,3], Xiang-Xiang Song[1,2,3]*, and Guo-Ping Guo[1,2,3,4]

1. CAS Key Laboratory of Quantum Information, University of Science and Technology of China, Hefei, Anhui 230026, China
2. Suzhou Institute for Advanced Research, University of Science and Technology of China, Suzhou, Jiangsu 215123, China
3. CAS Center for Excellence in Quantum Information and Quantum Physics, University of Science and Technology of China, Hefei, Anhui 230026, China
4. Origin Quantum Computing Company Limited, Hefei, Anhui 230088, China

† Y.-B. W. and Z.-Z. Z. contributed equally to this work.
* Author to whom correspondence should be addressed: X.-X. S. (songxx90@ustc.edu.cn)

# ABSTRACT

The excellent mechanical properties make graphene promising for realizing nanomechanical resonators with high resonant frequencies, large quality factors, strong nonlinearities, and the capability to effectively interface with various physical systems. Equipped with gate electrodes, it has been demonstrated that these exceptional device properties can be electrically manipulated, leading to a variety of nanomechanical/acoustic applications. Here, we review the recent progress of graphene nanomechanical resonators with a focus on their electrical tunability. First, we provide an overview of different graphene nanomechanical resonators, including their device structures, fabrication methods, and measurement setups. Then, the key mechanical properties of these devices, for example, resonant frequencies, nonlinearities, dissipations, and mode coupling mechanisms, are discussed, with their behaviors upon electrical gating being highlighted. After that, various potential classical/quantum applications based on these graphene nanomechanical resonators are reviewed. Finally, we briefly discuss challenges and opportunities in this field to offer future prospects of the ongoing studies on graphene nanomechanical resonators.

# 1. Introduction

Since its discovery twenty years ago in 2004[1], graphene, a single layer of carbon atoms arranged in a hexagonal lattice[2], has attracted significant research interest owing to its excellent electrical[3-6], optical[7-9] as well as mechanical[10-12] properties. In particular, its atomically thin feature[13], large Young's modulus[10, 14], and high breaking strength[10, 14] make graphene exceptional for mechanical applications. Meanwhile, graphene is electrically and optically accessible itself[15, 16], thereby can directly, and effectively, transduce the mechanical response into detectable electrical or optical signals[17]. With these advantages, graphene, along with its diverse two-dimensional (2D) material family, boosts the downscaling of the well-studied microelectromechanical systems (MEMS) into nanoelectromechanical systems (NEMS).

Clamping a suspended graphene membrane on a supporting substrate and making it vibrate, a simple NEMS device, a nanomechanical resonator, is realized[17]. Such graphene nanomechanical resonators exhibit high resonant frequencies[18-20], large quality factors[21-23], and possess sensitive responses upon external excitations[24, 25]. More importantly, the excellent electrical tunability provides extreme flexibility in *in-situ* manipulating the mechanical response of the device. This includes tuning the resonant frequency over a wide range[26, 27], engineering the paths of energy dissipation[21], changing the nonlinear effect from hardening to softening[22], and controlling the mode coupling between different vibrational modes[23, 27]. These characteristics of graphene nanomechanical resonators lead to a variety of fascinating applications, including mechanical switches[28, 29], acoustic microphones[30, 31], various sensors[24, 32-34], tunable phononic crystals[35, 36], and interfaces to quantum systems[18, 37, 38], demonstrating a promising future not only as next-generation NEMS but also for quantum applications.

In this review, we will start with an overview of device structures and experimental setups of various graphene nanomechanical resonators. Then, we will introduce the fundamental mechanical properties of these devices, in terms of their resonant frequencies, nonlinearities, dissipations, and mode coupling mechanisms. The excellent electrical tunability of these properties will be particularly highlighted. Finally, we will review potential classical/quantum mechanical applications based on graphene nanomechanical resonators, followed by a brief outlook on future challenges and opportunities. We would like to note that although graphene is referred to as a 2D material that consists of only one layer of carbon atoms, we will also include those devices made of multiple layers in this review and label them as bilayer/trilayer/few-layer graphene devices. Besides, we would like to list several related reviews[39-45] which can be referenced. For example, a comprehensive review on the mesoscopic physics of nanomechanical systems has been given in Ref. [39]. For nanomechanical resonators based on different 2D materials, see Refs. [40, 41]. The dynamics of these 2D material membranes are reviewed in Refs. [42, 45]. From the view of applications, Ref. [43] reviews various sensors based on nanomechanical resonators made of 2D materials, while Ref. [44] concentrating on graphene-based devices in MEMS and NEMS applications. In this review, we will mainly focus on the recent progress of graphene nanomechanical resonators and the underlying physical mechanisms.

**2. Devices and experimental setups**

Various types of graphene nanomechanical resonators have been investigated so far. Figure 1 shows some typical examples.

The most widely-studied devices are doubly clamped[17-19, 25, 33, 46-51] (Fig. 1a[19]) and fully clamped[32, 52-58] (Fig. 1b[57]) resonators. Their difference in the clamping conditions results in distinct stress distributions and mode shapes. Compared with doubly clamped resonators in which two edges are freely vibrating, fully clamped devices are expected to have more regular and predictable vibrational modes[50, 59]. However, in order to make a gate electrode sitting underneath the suspended graphene membrane for electrical tuning, doubly clamped devices have the advantage in fabrication flexibility. In addition to using a heavily doped substrate as the global back gate[17, 19, 33, 46, 50], a local back gate (which is beneficial for electrical readout and will be discussed later) to the doubly clamped resonator is easier to fabricate by depositing metals directly inside the trench via lithography[18, 27] or through the self-alignment technique[23, 25]. However, in order to maintain the fully clamped boundary, a buried local gate electrode is needed, so that it can be connected to outer metal pads for applying voltages[52], which makes the fabrication process more complicated.

Alternative devices have been designed to include two narrow channels[21, 22, 24, 60-62] (Fig. 1c[61]). This structure allows directly depositing metals to fabricate a local electrode while keeping most parts of the boundary clamped. The two narrow channels also act as venting channels to connect the outer environment to the sealed area underneath the suspended circular graphene membrane. Another example is to use an additional layer of SU-8 polymer to define a fully clamped resonator while electrodes are arranged in a configuration similar to that of doubly clamped devices[26, 63] (Fig. 1d[26]). There are also other device designs, such as trampoline-shape[64] (Fig. 1e) or H-shape[65] (Fig. 1f). If the graphene flake is sufficiently thick, it even can serve as a singly clamped resonator[66] (Fig. 1g).

Moreover, graphene nanomechanical resonators can be extended to couple with each other. Figure 1h shows a device consisting of three doubly clamped resonators which are coupled in series[23]. Similarly, a dumbbell-shaped resonator is studied (Fig. 1i), in which two circular mechanical cavities are connected by the venting channel in between[67] so that their vibrational modes can interact. A large array of graphene resonators is also available[68, 69] (Fig. 1j[68]). If mode couplings are simultaneously generated, phononic crystals can be expected. Two approaches have been demonstrated so far. One is to selectively shape a suspended graphene membrane to form a periodic pattern[35] (Fig. 1k). The other is to transfer a graphene flake to a pre-patterned standing pillar array[36] (Fig. 1l). Both devices are equipped with a gate electrode for electrical tuning, respectively.

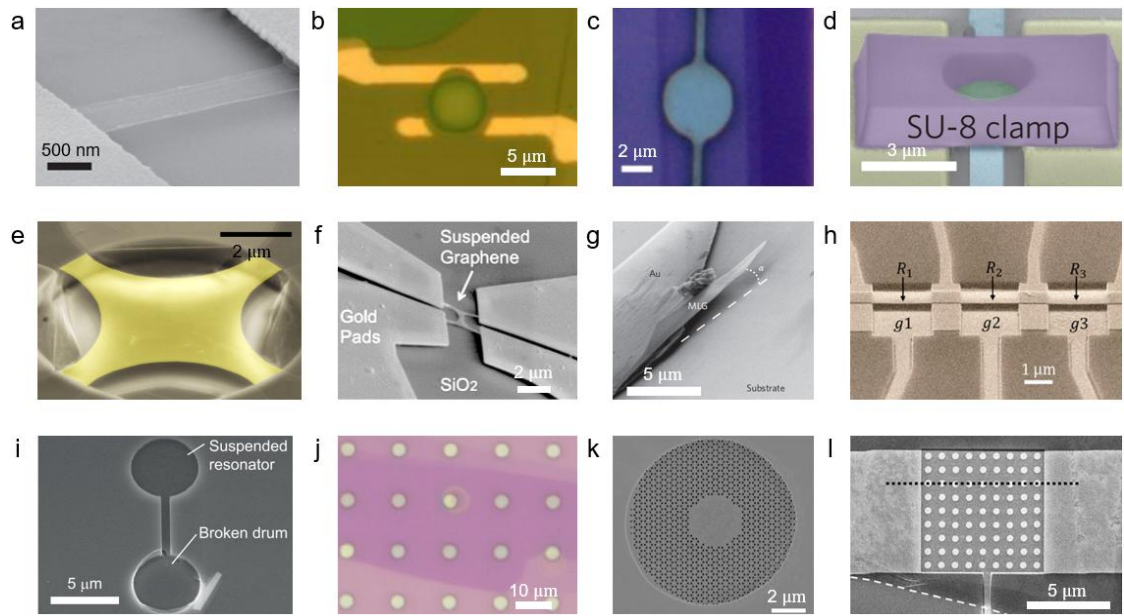

**Figure 1. Various graphene nanomechanical resonators.** (a) Doubly clamped resonator. (b) Fully clamped resonator. (c) Fully clamped resonator with venting channels. (d) Fully clamped resonator using an additional layer of SU-8 polymer. (e) Trampoline-shaped resonator. (f) H-shaped resonator. (g) Singly clamped resonator. (h) Three doubly clamped resonators coupled in series. (i) Dumbbell-shaped resonator with a venting channel in between. (j) Large array of drum resonators. (k) Phononic crystal by shaping a suspended graphene membrane into a periodic pattern. (l) Phononic crystal by transferring a graphene flake to a pre-patterned standing pillar array. (a) Reproduced with permission.[19] Copyright 2011, Springer Nature. (b) Reproduced with permission.[57] Copyright 2018, American Chemical Society. (c) Reproduced under the terms of a creative commons CC-BY international license.[61] Copyright 2020, The Authors, published by Springer Nature. (d) Reproduced with permission.[26] Copyright 2013, Springer Nature. (e) Reproduced under the terms of a creative commons CC-BY international license.[64] Copyright 2019, The Authors, published by Springer Nature. (f) Reproduced with permission.[65] Copyright 2015, American Chemical Society. (g) Reproduced with permission.[66] Copyright 2012, Springer Nature. (h) Reproduced under the terms of a creative commons CC-BY international license.[23] Copyright 2019, The Authors, published by National Academy of Sciences of America. (i) Reproduced under the terms of a creative commons CC-BY-NC-ND international license.[67] Copyright 2021, The Authors, published by American Chemical Society. (j) Reproduced with permission.[68] Copyright 2011, Springer Nature. (k) Reproduced under the terms of a creative commons CC-BY international license.[35] Copyright 2021, The Authors, published by American Chemical Society. (l) Reproduced with permission.[36] Copyright 2021, American Chemical Society.

Although the detailed structures of graphene nanomechanical resonators are different from each other, their fabrication methods can be roughly divided into two categories. Their main difference lies in the time of transferring graphene onto the substrate. In the first category[19, 26, 46, 66, 70, 71], the graphene flake is either mechanically exfoliated or transferred onto a flat substrate first. After shaping the flake and defining the electrodes, the substrate underneath the graphene flake is etched. Finally, hot acetone or critical point dry are employed to suppress the influence of the sudden change in surface tension when taking the device out of the liquid etchant[5, 46]. This approach enables the fabrication of scalable devices with designed shapes based on exfoliated[19, 46, 66] or large-area chemical vapor deposited (CVD) graphene membranes[70]. The drawbacks are 1) the wet etching process may leave significant residues on the surface of the graphene flake; and 2) the difficulty in fabricating a local back gate since it needs to be buried before transferring and etching. In the second category[17, 18, 26, 38, 49, 52], the substrate is prepatterned to form trenches or holes first. After defining the electrodes, the graphene flake is finally exfoliated or transferred onto the pre-patterned substrate. The difference in the final step is the polymer-assisted transfer technique[72, 73] significantly enhances the yield and the scalability at the cost of inducing potential polymer residues, while mechanical exfoliation results in clean flakes which are randomly distributed on the substrate and only some of them may be suspended over a trench or a hole[69, 74]. The second approach enables more flexibility in patterning the substrate and the electrodes so that it is easier to integrate local back gates[18, 38, 52, 75]. However, transferring/exfoliating the graphene flake and maintaining its suspension on a pre-patterned substrate may bring uncertainty, especially for large-scale integration.

There are a variety of measurement schemes for actuating, transducing, and detecting the mechanical vibration of graphene nanomechanical resonators. In this review, we will briefly introduce them. For more detailed discussions on this topic, see Refs. [41-43].

The optical measurement method is widely used for investigating graphene nanomechanical resonators[32, 54, 55, 58, 67, 70, 71, 76, 77]. A common setup is to focus a frequency-modulated laser beam on the graphene membrane. This causes graphene to expand and contract due to the thermal effect, actuating vibrations. Another laser beam, with a much lower power, is applied for optical detection. When the detection laser shines on the graphene membrane, part of it reflects while the rest passes through the membrane and is reflected by the substrate, generating interference patterns. The intensity of the reflected signal is highly sensitive to the position of the membrane. By modulating the intensity of the reflected signal through a photodiode, the mechanical vibration of the graphene can be detected.

Electrical actuation[48, 49, 52, 63] is achieved by applying a DC gate voltage $V_{\text{DC}}$ and an AC microwave $V_{\text{AC}}$ between the gate electrode and the graphene membrane, generating an electrostatic force $F_{\text{DC}}$ that pulls the membrane towards the gate:

$$F_{\text{DC}} = \frac{1}{2}\frac{\partial C_g}{\partial z}V_{\text{DC}}^2, \quad (2.1)$$

and an oscillating force $F_{\text{AC}}$ that drives the membrane to vibrate:

$$F_{\text{AC}} = \frac{\partial C_g}{\partial z}V_{\text{DC}}V_{\text{AC}}, \quad (2.2)$$

where $C_g$ is the capacitance between the gate electrode and the membrane, and $z$ is the displacement along the vertical direction where the membrane vibrates. The vibration is then transduced to the current change to be detected. In graphene nanomechanical resonators, the transducing mechanism is mainly based on the field effect[48, 52], rather than the piezoresistive effect[78, 79]. Namely, the mechanical vibration modulates the capacitance between the gate and the membrane, contributing to an additional current through a relatively large transconductance.

Experimentally, directly measuring such a current at the resonant frequency $f_0$ (typically tens of MHz) is challenging. This is because the parasitic capacitance and the contact resistance act as a low-pass filter, inducing signal leakage[19, 46]. To suppress this effect, replacing the global back gate with a local gate electrode underneath the suspended membrane to reduce the parasitic capacitance is essential[18, 26, 52]. Meanwhile, frequency mixing techniques are employed. Such techniques mix the mechanical vibration signal at $f_0$ with a signal at $f_0 + \Delta f$ to generate a down-converted signal at the frequency of $\Delta f \sim$ kHz to be detected. The most widely-used frequency mixing techniques include two-source mixing[27, 46, 51, 80], amplitude modulation[81, 82], and frequency modulation techniques[19, 20, 23]. Here, we would like to point out that, compared with devices based on other 2D materials, such as $MoS_2$, it is much easier to achieve high mobility and low contact resistance in graphene-based devices, especially at low temperature. These features are beneficial for effectively transducing the mechanical vibrations to electrically detectable signals. An important point of the electrical measurement method is to faithfully separate the electrical contribution of the device from the transduced mechanical contribution[83, 84].

In addition to pure optical and electrical measurement methods, combined strategies are also available, such as electrical driving and optical detecting the mechanical response[17, 53, 61, 85]. Moreover, probing the mechanical vibration of graphene membranes using an integrated detector (such as a microwave cavity) is also demonstrated[22, 24, 38, 75, 86-88]. Other strategy such as directly monitoring the mechanical vibration using a scanning probe microscope is also available[50]. Typical measurement schemes and their characteristics are briefly summarized in Table 1.

| Measurement Schemes | Mechanisms | Reported working temperature | Characteristics |
|---|---|---|---|
| Optical | Interference | ~4 K-1200 K | Requires sufficiently large suspension area; Readout responsivity is sensitive to suspension height; Able to resolve mode shape |
| Electrical | Field effect; Piezoresistive effect; Capacitance modification; | ~20 mK-300 K | Requires sufficiently large transconductance/piezoresistive gauge factor; Requires separating transduced mechanical contribution from electrical contribution |
| Via integrated detectors | Optomechanical coupling | ~20 mK-300 K | No need to directly drive resonator on resonance; Requires sufficiently large coupling strength; |
| Scanning probe microscopy | Atomic force | 300 K | Limited bandwidth for high frequency applications; Difficulty in extending working temperature range; Able to resolve mode shape |

**Table 1. Typical measurement schemes for graphene nanomechanical resonators.**

## 3. Properties of tunable graphene nanomechanical resonators

### 3.1 Single resonator

The most important property of a mechanical resonator is its resonant frequency. It strongly depends on various parameters, such as the thickness of the graphene flake, the geometry of the resonator, the clamping condition, and the initial built-in strain. When the suspended graphene flake consists of only few layers, the influence of bending rigidity on the resonator can be neglected[89, 90]. Thus, the mechanical response of the resonator can be described as a tensioned membrane. In such a regime, the resonant frequency of a doubly clamped resonator is given as[81]:

$$f_{\text{doubly clamped}}^{\text{membrane}} = \frac{1}{2L}\sqrt{\frac{E}{\rho}\varepsilon_0} = \frac{1}{2L}\sqrt{\frac{\sigma_0}{\rho t N}}, \quad (3.1)$$

where $L$ is the suspending length of the membrane, $E \approx 1\,\text{TPa}$ is the Young's modulus of graphene[10], $\rho \approx 0.74\,\text{mg/m}^2$ is the mass density, $\varepsilon_0$ is the built-in strain in the membrane, $\sigma_0$ is tensile stress, $t = 0.34\,\text{nm}$ is the interlayer distance of graphene, and $N$ is the number of graphene layers. Similarly, the resonant frequency of a circular fully clamped resonator is given as[81]:

$$f_{\text{fully clamped}}^{\text{membrane}} = \frac{0.766}{D}\sqrt{\frac{E}{\rho}\varepsilon_0} = \frac{0.766}{D}\sqrt{\frac{\sigma_0}{\rho t N}}, \quad (3.2)$$

where $D$ is the diameter of the suspended membrane. It can be seen that the resonant frequency is modified by different clamping boundary conditions while scaling in a similar way of geometric size and built-in strain. Note that the resonant frequency is proportional to $\sqrt{\frac{E}{\rho}}$. The large $E$ and low $\rho$ of graphene, compared with other 2D materials[43], tend to result in nanomechanical resonators with higher frequencies.

Figure 2a shows calculated resonant frequencies as a function of geometric size (length/diameter for doubly/fully clamped resonators, respectively) under different built-in strains based on the parameters of graphene. The resonant frequency increases as lowering the geometric size or increasing the strain. Typical experimental parameters lie in the range of several micrometers in the geometric size and ~0.05% in the built-in strain, resulting in tens of MHz in the frequency (roughly in the blue-shaded area).

In the limit of thick flakes, their mechanical responses change from membrane-like to plate-like behaviors. For example, for fully clamped resonators, in the plate-like regime, the resonant frequency is given as[91]:

$$f_{\text{fully clamped}}^{\text{plate}} = \frac{3.22}{D^2}\sqrt{\frac{Yt^2N^2}{3\rho(1-\nu^2)}}, \quad (3.3)$$

where $Y = Et$ and $\nu = 0.165$ is the Poisson ratio[81]. Comparing Eq. (3.3) with Eq. (3.2), the frequency is now determined by the stiffness, and its geometric scaling behavior is changed.

Such an interesting crossover between membrane-like and plate-like behaviors has been systematically investigated in other 2D nanomechanical resonators[92-96]. Simply assuming a combined influence on the resonant frequency $f_0$ as[45]:

$$f_0 = \sqrt{(f_{\text{fully clamped}}^{\text{membrane}})^2 + (f_{\text{fully clamped}}^{\text{plate}})^2}, \qquad (3.4)$$

the experimental results can be well understood. In a similar way, we calculate the resonant frequency crossover according to Eq. (3.4) based on the parameters of graphene, as shown in Fig. 2b. It can be seen that with the increased thickness, the membrane-like frequency decreases and the plate-like frequency increases. A larger geometric size or larger initial stress tends to result in a crossover occurring at a larger thickness. We further benchmark several typical experimental data points[22, 54, 56, 61, 97, 98] reported in the literature with the calculated curves, and find a reasonable agreement similar to that observed in nanomechanical resonators made of WSe$_2$[94], hBN[95], and MoS$_2$[92, 93, 96].

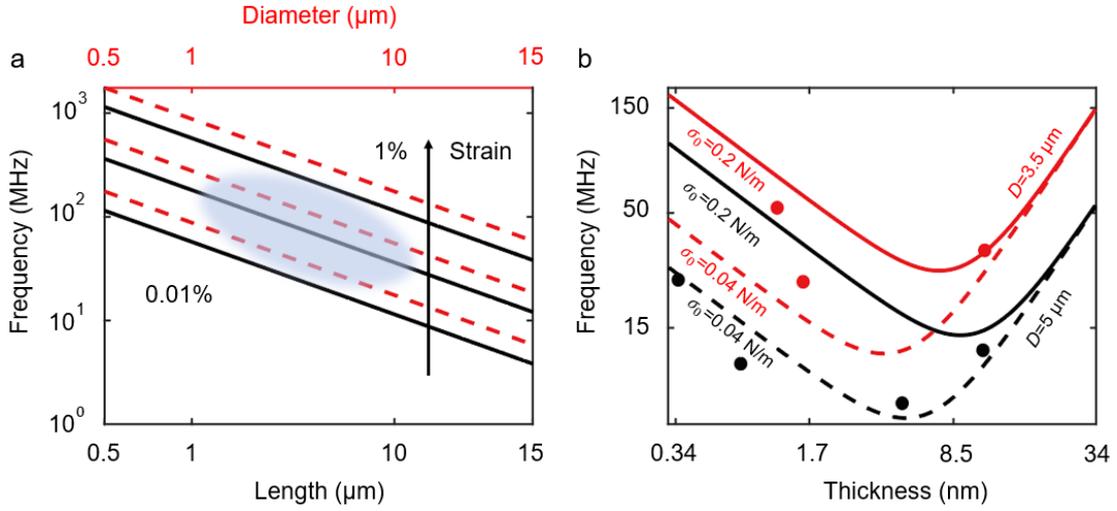

**Figure 2. Resonant frequency scaling.** (a) Calculated resonant frequencies (black solid/red dashed lines for doubly/fully clamped resonators, respectively) in the membrane-like regime as a function of geometric size (length/diameter for doubly/fully clamped resonators, respectively) under different built-in strains based on parameters of graphene. (b) Calculated resonant frequencies (curves) of fully clamped graphene resonators as a function of flake thickness, where contributions of membrane-like and plate-like behaviors are considered. Black (red) curves represent devices with the diameter of 5 μm (3.5 μm). The solid (dashed) curves represent the calculated results under tensile stress $\sigma_0$= 0.2 N/m (0.04 N/m). Data points in different colors are experimental results obtained from devices with corresponding geometric sizes, which are extracted from Refs. [54, 56, 61, 97] (black) and Refs. [22, 96] (red), respectively.

One of the main features of graphene nanomechanical resonators is the strong electrical tuning effect on the resonant frequency. Here, we introduce its physical mechanism. The resonant frequency $f_0$ can be written as:

$$f_0 = \frac{1}{2\pi}\sqrt{\frac{k}{m_{\text{eff}}}}, \qquad (3.5)$$

where $m_{\text{eff}}$ is the effective mass of the vibrational mode. $k$ is the spring constant that can be calculated as:

$$k = \frac{\partial^2 U_{\text{total}}}{\partial z^2}, \quad (3.6)$$

where $U_{\text{total}} = U_{\text{elastic}} + U_{\text{electrostatic}}$ is the total energy with $U_{\text{elastic}}$ ($U_{\text{electrostatic}}$) as the elastic (electrostatic) energy.

Taking the example of doubly clamped resonators, according to a continuum mechanics model, we have[99, 100]:

$$U_{\text{elastic}} = \frac{8ES\varepsilon_0 z^2}{3L} + \frac{64ESz^4}{9L^3}, \quad (3.7)$$

$$U_{\text{electrostatic}} = -\frac{1}{2}C_g V_g^2, \quad (3.8)$$

where $S$ is the cross-sectional area of the graphene membrane, $V_g$ is the applied DC gate voltage. We can define:

$$k_1 = \frac{\partial^2 U_{\text{elastic}}}{\partial z^2} = \frac{16ES\varepsilon_0}{3L} + \frac{256ESz^2}{3L^3}, \quad (3.9)$$

$$k_2 = \frac{\partial^2 U_{\text{electrostatic}}}{\partial z^2} = -\frac{1}{2}\frac{d^2 C_g}{dz^2}V_g^2. \quad (3.10)$$

Note that in $k_1$, $z$ represents the maximum vertical displacement at the center of the graphene membrane, which can be obtained by solving the equilibrium condition of $\partial U_{\text{total}}/\partial z = 0$:

$$\frac{256ES}{9L^3}z^3 + \left(\frac{16ES\varepsilon_0}{3L} - \frac{1}{2}\frac{d^2 C_g}{dz^2}V_g^2\right)z - \frac{1}{2}\frac{dC_g}{dz}V_g^2 = 0. \quad (3.11)$$

Physically, with the increasing of $V_g$, the suspended membrane is pulled towards the gate electrode, resulting in a larger displacement $z$. Therefore, the second term in $k_1$ leads to a hardening effect due to the stretching. However, $\frac{d^2 C_g}{dz^2}$ in Eq. (3.10) is usually positive (for example, considering a parallel-plane capacitor model) so that $k_2$ contributes to a capacitive softening effect to lower the frequency.

Finally, Eq. (3.5) can be expressed as:

$$f_0 = \frac{1}{2\pi}\sqrt{\frac{1}{m_{\text{eff}}}\left(\frac{16ES\varepsilon_0}{3L} + \frac{256ESz^2}{3L^3} - \frac{1}{2}\frac{d^2 C_g}{dz^2}V_g^2\right)}. \quad (3.12)$$

It can be found that the 2D nature of graphene (the thickness is much smaller than the suspending length) results in a relatively small spring constant[45]. As a consequence, a small out-of-plane electrostatic force can generate a significant membrane bending, compared with the thickness of graphene itself, leading to a large electric tuning range of the resonant frequency. From Eq. (3.12), we find that the frequency tuning is symmetric with respect to $V_g$. When $V_g = 0$, $f_0$ is dominated by the built-in strain $\varepsilon_0$. In addition, $\varepsilon_0$ also modulates the frequency tuning behaviors. Figure 3a-c shows three typical frequency tuning curves calculated using Eq. (3.12), under low, medium and high $\varepsilon_0$, respectively. Similar experimental results have been observed[38, 46, 64], as shown in Fig. 3d-f, respectively.

For fully clamped resonators, we have similar results[100, 101]:

$$k = \frac{2\pi E t \varepsilon_0}{1-v^2} + \frac{32\pi E t z^2}{(1-v^2)D^2} - \frac{1}{2}\frac{d^2 C_g}{dz^2}V_g^2, \quad (3.13)$$

in which $z$ is determined by:

$$\frac{32\pi E t}{3(1-v^2)D^2}z^3 + \left(\frac{2\pi E t \varepsilon_0}{1-v^2} - \frac{1}{2}\frac{d^2 C_g}{dz^2}V_g^2\right)z - \frac{1}{2}\frac{dC_g}{dz}V_g^2 = 0. \quad (3.14)$$

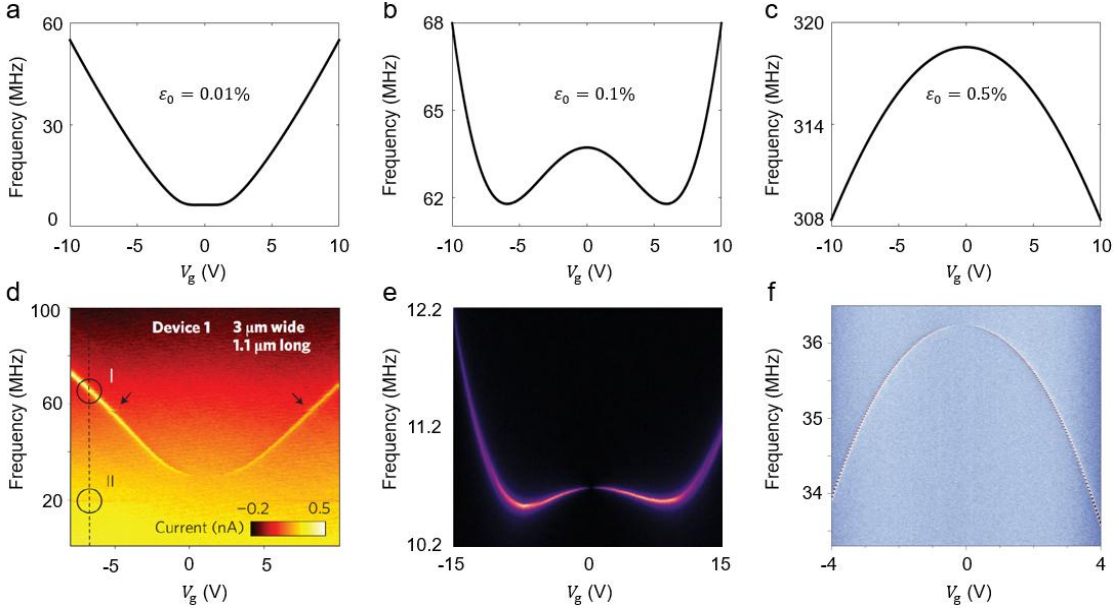

**Figure 3. Electrical tuning of resonant frequency.** (a)-(c) Calculated resonant frequency tuning behaviors using Eq. (3.12) with different built-in strains $\varepsilon_0$= 0.01 %, 0.1 %, and 0.5 %, respectively. (d)-(f) Corresponding experimental frequency tuning spectra. (d) Reproduced with permission.[46] Copyright 2009, Springer Nature. (e) Reproduced under the terms of a creative commons CC-BY international license.[64] Copyright 2019, The Authors, published by Springer Nature. (f) Reproduced with permission.[38] Copyright 2014, Springer Nature.

Next, we will discuss the nonlinearity of the mechanical vibration. The dynamics of a graphene nanomechanical resonator can be modeled as:

$$\ddot{z} + \omega_0^2 z + \gamma \dot{z} = F_{\text{drive}}/m_{\text{eff}}, \quad (3.15)$$

where $\omega_0 = 2\pi f_0$ is the resonant angular frequency, $\gamma$ is the damping rate, and $F_{\text{drive}}$ is the driving force. Equation (3.15) holds at small vibrational amplitude $z$. When $F_{\text{drive}}$ increases, the influence of nonlinearity emerges[53, 61, 102]. Namely, the restoring force $F_{\text{res}}$ cannot be simply modeled as:

$$F_{\text{res}} = kz = m_{\text{eff}}\omega_0^2 z. \quad (3.16)$$

We can expand $F_{\text{res}}$ to higher orders to account for the nonlinearity:

$$F_{\text{res}} = k_1 z + k_2 z^2 + k_3 z^3 + k_3 z^4 + \cdots. \quad (3.17)$$

Unless broken[103-105], there is a symmetry that $F_{\text{res}}(z) = -F_{\text{res}}(-z)$ so that only odd terms need to be considered. Therefore, the leading term in nonlinearity has the form of $\alpha z^3$, which is known as the Duffing nonlinearity[53, 104-109].

The Duffing nonlinearity shifts the resonant frequency upward or downward depending on the sign of $\alpha$. As shown in Fig. 4a, the amplitude-frequency response deviates from the Lorentzian shape in the linear regime, exhibiting a backbone shape with a positive value of $\alpha$ (dashed curve)[76]. This results in a region, within which three branches with different amplitudes emerge under a given driving frequency. Among them, the two with the highest and the lowest amplitudes are stable. Experimentally, sudden switches between the high-amplitude and the low-amplitude branches are observed. More interestingly, the position of the switch in frequency is related to the sweeping direction of the external driving frequency (solid blue and yellow curves), leading to a frequency hysteresis[19, 61]. The bistable feature is more clearly demonstrated by fixing the driving frequency within this region and applying a noise signal to the gate electrode. As shown in Fig. 4b, the time-resolved measurement demonstrates stochastic switches between the two branches[76]. Besides, the Duffing nonlinearity has also been revealed by the ring-down experiment. As shown in Fig. 4c, a graphene resonator is firstly driven into the nonlinear regime. At $t = 0$, the driving signal is turned off and the frequency evolution is monitored. As the vibrational amplitude of the resonator gradually decreases due to the dissipation, the resonant frequency also shifts, indicating the influence of the Duffing nonlinearity[21].

In graphene nanomechanical resonators, the apparent Duffing coefficient $\alpha$ arises from two contributions[104-106]: 1) mechanical nonlinearity due to displacement-dependent tension (determined by the geometric parameters of the device if the symmetry in $F_{\text{res}}(z)$ is preserved), and 2) capacitive nonlinearity since the capacitance is a nonlinear function of the displacement (can be tuned by gate voltages via derivatives of the capacitances). Therefore, upon tuning the gate voltage, the competition between these two origins can be manipulated. As shown in Fig. 4d, the apparent $\alpha$ is tuned from positive (leftmost panel) to negative (rightmost panel) through electrical gating, demonstrating a transition from hardening to softening[22]. In particular, $\alpha$ can be adjusted to near zero, so that the influence of higher-order nonlinearities[47] can be better resolved experimentally[104].

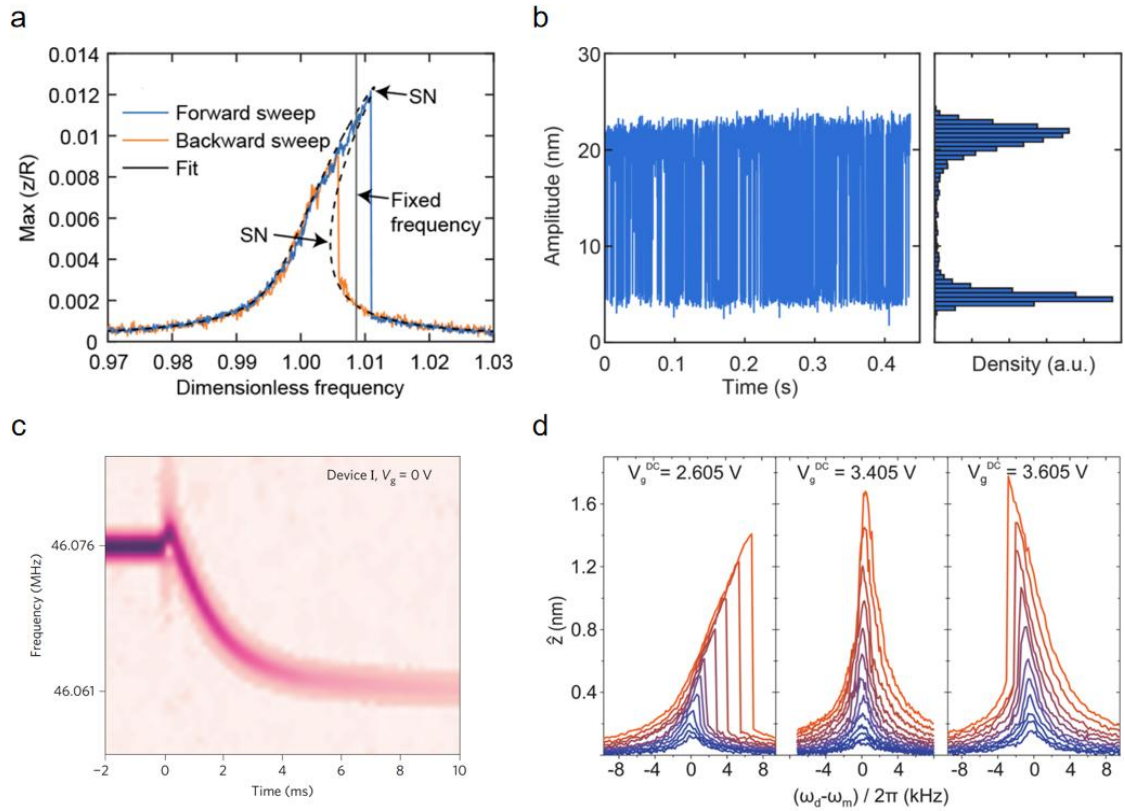

**Figure 4. Duffing nonlinearity in graphene nanomechanical resonators.** (a) Theoretical (black dashed) and experimental (blue and yellow solid) curves of amplitude-frequency response under strong external driving. Experimental results exhibit a frequency hysteresis under forward (blue) and backward (yellow) sweeping directions. (b) Stochastic switches between the high-amplitude and low-amplitude branches at the fixed driving frequency shown in (a). (c) Time-resolved measurement demonstrates a frequency shift when the vibrational amplitude decreases. (d) Gate tuning of the apparent Duffing coefficient from positive (leftmost panel) to negative (rightmost panel). (a)-(b) Reproduced under the terms of a creative commons CC-BY-NC-ND international license.[76] Copyright 2019, The Authors, published by American Chemical Society. (c) Reproduced with permission.[21] Copyright 2017, Springer Nature. (d) Reproduced with permission.[22] Copyright 2014, American Chemical Society.

In Eq. (3.15), $\gamma$ describes the rate at which the energy stored in the resonator dissipates. It can be extracted from the full-width-at-half-maximum (FWHM) of the resonance peak in the frequency domain[19, 22-24] or from the time-resolved ring-down measurement[21, 85]. Its value is usually evaluated by the quality factor as $Q = \omega_0/\gamma$. Although such $Q$ values have been widely measured, the dominating physical origins of the energy dissipation in graphene nanomechanical resonators are still far from fully understood[110].

Table 2 benchmarks key parameters of typical graphene nanomechanical resonators reported in the literature. It can be found that $Q$ strongly depends on the temperature[46, 70, 111, 112] with typical values of 10-1500 at room temperature and significantly increase to $> 10^5$ at low temperature. However, its dependences on the geometric parameters of the device and thicknesses of the membrane are much less obvious[17, 55], probably due to different strain distributions accumulated during fabrication processes among different groups. It has been suggested that various extrinsic factors (such as viscous dissipation, clamping losses, Ohmic losses, optomechanical coupling, coupling to two-level systems, adhesion to the substrate) and intrinsic factors (such as thermoelastic or Akhiezer dissipation, dissipation due to defects or interlayer friction) may contribute to the energy dissipation[60, 97, 110, 113, 114]. However, neither of them can fully explain the experimental results in terms of both the scaling behaviors and the order of magnitude in dissipation. Moreover, in addition to the commonly observed linear damping $\gamma \dot{z}$, nonlinear damping that scales with the vibrational amplitude ($\eta z^2 \dot{z}$)[115, 116] has been identified to play an important role in graphene nanomechanical resonators[19, 54, 87, 117]. Therefore, unraveling the dominating mechanisms underlying the energy dissipation in graphene nanomechanical resonators is challenging, yet very important, and is still ongoing.

| Geometry structure | Working temperature | Quality factor | Resonant frequency (MHz) | Geometric size (μm) | Thickness |
|---|---|---|---|---|---|
| Doubly clamped[50] | RT | 5 | 31 | 2.8 | 11 nm |
| Doubly clamped[70] | RT | 52/44 | 9.77/19.2 | 2 | MLG |
| Doubly clamped[33] | RT | 125/-/- | 70/26/87 | 1.6/1.3-1.6/1.3-1.6 | MLG |
| Doubly clamped[17] | RT/RT/RT/50 K | 210/78/60/1800 | 42/70.5/35.8/- | 5/1.1/2.7/- | 15 nm/MLG/5 nm/20 nm |
| Doubly clamped[71] | RT | 97/704 | 8.36/19.82 | 8/- | 1 nm |
| Doubly clamped[46] | RT/RT/5 K | 125/-/14000 | 65/42/130.1 | 1.1/1.8/1.4 | MLG |
| Doubly clamped[49] | 77 K | 10000 | 34 | 2 | MLG |
| Doubly clamped[26] | 77 K | - | 50.7 | 4.2 | MLG |
| Doubly clamped[86] | 4.2 K | 1400/-/- | 57/178/73 | 1.5/0.7/1 | MLG/MLG/FLG |
| Doubly clamped[18] | 4.3 K | 4700 | 123 | 2.4 | MLG |
| Doubly clamped[48] | 270 mK | 33447 | 83.618 | 2 | 5 layers |
| Doubly clamped[19] | 90 mK/4 K | 100000/1500 | 156/200 | 2/1.7 | MLG |
| Doubly clamped[75] | 22 mK | 15000 | 24 | 2.5 | BLG |
| Doubly clamped[27] | 10 mK | 3500/3000/3000 | 98.05/105/101 | 2 | 5 layers |
| Doubly clamped[23] | 10 mK | 111000/115000/196000 | 133/134/144 | 2 | 7 layers |
| Fully clamped[57] | 450 K/800 K/1200 K | 750/300/30 | 100/40/30 | 3/5/4 | MLG/TLG/BLG |
| Fully clamped[32] | RT | 20/- | 10.9/26 | 10/5 | 7 nm/BLG |
| Fully clamped[58] | RT | 25 | 66 | 4.75 | MLG |
| Fully clamped[56] | RT | 42/53/32 | 13.6/16/38.5 | 5 | 4 nm/5 nm/18 nm |
| Fully clamped[59] | RT | 50/40/20 | 14.1/24.3/61.5 | 5 | 5 nm/4.5 nm/MLG |
| Fully clamped[26] | RT | 55/15/-/84 | 52.19/25.04/100/35 | 4/2/3/3 | MLG |
| Fully clamped[63] | RT | 60/54 | 48/260 | 4/1.5 | MLG |
| Fully clamped[118] | RT | 57/67 | 8.6/3.0 | 8/20 | MLG |
| Fully clamped[53] | RT | 138/74 | 14.7/28 | 5/4 | 5 nm/8 nm |
| Fully clamped[54] | RT | 454/- | 20.1/18 | 5/5 | 10 nm/14 nm |
| Fully clamped[55] | RT | 2400 | 3.65 | 22.5 | MLG |
| Fully clamped[111] | 20 K | 1000 | 70 | 2 | TLG |

| Device type | Temperature | Frequency (MHz) | Q factor | Size (μm) | Thickness |
|---|---|---|---|---|---|
| Fully clamped[52] | 5 K | 1066/691 | 94.9/47 | 3.5 | 3.5 nm |
| Fully clamped with vents[98] | RT | 42/-/-/- | 18.2/25/10/11 | 5 | BLG/MLG/BLG/FLG |
| Fully clamped with vents[60] | RT | 57/53.2/40/31 | 9.2/21/14/19 | 5 | BLG/MLG/ BLG/ BLG |
| Fully clamped with vents[62] | RT | 500/- | 3.46/5.2 | 14/5.5 | MLG |
| Fully clamped with vents[61] | RT | 1700/-/2200 | 33.8/27.65/39.2 | 6/6/- | MLG |
| Fully clamped with vents[22] | 30 mK | 100000/17700 | 57.5/33.3 | 3.5 | TLG/4 layers |
| Fully clamped with vents[24] | 15 mK | -/200000 | 67/46 | 3.2 | 25/5-6 layers |
| Fully clamped with vents[21] | 15 mK | 1000000/-/- | 44/44/67 | 3.3 | 5-6/35/25 layers |
| Fully clamped with vents[38] | 14 mK | 220000 | 36.3 | 4 | 10 nm |
| Dumbbell-shaped[31] | RT | 4/5 | 28/51 | 10 | 27 nm/6 nm |
| Dumbbell-shaped[119] | RT | 100 | 13 | 5 | 10.5 nm |
| Dumbbell-shaped[67] | RT | 107 | 15.9 | 5 | MLG |
| Dumbbell-shaped[76] | RT | 416.6 | 13.92 | 5 | MLG |
| Trampoline-shaped[64] | RT | 910/4400/1400 | 10.7/25.4/21.9 | 8/6/6 | MLG |
| H-shaped[65] | RT | 1000 | 1.191 | 1.8 | MLG |
| Singly clamped[66] | RT | 26.1 | 1.2 | - | 30 nm |

**Table 2. List of key parameters of typical graphene nanomechanical resonators reported in the literature.** "RT" stands for room temperature. "MLG", "BLG", "TLG", and "FLG" represent monolayer, bilayer, trilayer, and few-layer graphene, respectively. "-" means the corresponding parameter is not provided. The geometric sizes for doubly clamped, trampoline-shaped, and H-shaped devices are extracted as the suspending length between the electrical contacts, while are extracted as the diameter of the suspended circular region for fully clamped, fully clamped with vents, and dumbbell-shaped devices.

**3.2 Coupled system**

A nanomechanical resonator has different vibrational modes[50, 52, 54, 59, 118] which can couple with each other. Moreover, different resonators can interact with each other as well, forming coupled resonator arrays[23, 27]. In addition to frequency tuning, the excellent electrical tunability of graphene nanomechanical resonators also offers the powerful capability of coupling engineering.

We first consider the linear coupling. The potential energy of two linearly coupled harmonic modes is written as[39]:

$$U_{12} = \frac{1}{2}m_1\omega_1^2 x_1^2 + \frac{1}{2}m_2\omega_2^2 x_2^2 + m_0 g_{12} x_1 x_2, \quad (3.18)$$

where $m_i$, $\omega_i$, and $x_i$ are the effective mass, the resonant angular frequency, and the displacement of the mode $i$ ($i$=1, 2), respectively. $m_0 g_{12}$ characterizes the linear coupling strength. When tuning, for example, $\omega_1$ to approach $\omega_2$ by gate voltage $V_{g1}$, an avoided crossing is expected, as shown in Fig. 5a. At the avoided crossing regime, the two eigenmodes (with frequencies of $\omega_\pm$, respectively) are hybridizations of original modes 1 and 2 (with frequencies of $\omega_1$ and $\omega_2$, respectively), with their splitting $|\omega_+ - \omega_-|$ determined by $g_{12}$ [120, 121].

In coupled graphene nanomechanical resonators, the linear mode coupling strength can be as large as ~10 MHz, which is ~10% of the resonant frequency (inset of Fig. 5a) and is significantly larger than the linewidth[23]. Such a strong coupling suggests that the energy can be efficiently transferred between the coupled resonators. This also allows the integration of more resonators to engineer mode coupling. Figure 5b shows a schematic illustration of three resonators coupled in series as $R_1$, $R_2$, and $R_3$. The direct neighboring couplings $\Omega_{12}$ ($\Omega_{23}$) between $R_1$ ($R_3$) and $R_2$ generate an indirect coupling $\Omega_{13}$ between spatially-separated $R_1$ and $R_3$ [27]. When the frequencies of $R_1$ and $R_3$ are adjusted to be degenerated ($\omega_1 = \omega_3$), while are detuned from the frequency of $R_2$ by a value of $\Delta_{12} = \Delta_{23} = |\omega_1 - \omega_2| = |\omega_2 - \omega_3| = \Delta$, $\Omega_{13}$ has a simple form as[27]:

$$\Omega_{13} = \frac{\Omega_{12}\Omega_{23}}{2\Delta}. \quad (3.19)$$

Therefore, the effective coupling strength $\Omega_{13}$ can be electrically manipulated by varying the frequencies of the resonators, thus $\Delta$.

Figure 5c shows the experimental demonstration of such a mode coupling engineering. As highlighted by the blue dotted circle, an indirect coupling $\Omega_{13}$ is generated and is modulated by varying $\Delta$ (Fig. 5d)[27]. As shown in Fig. 5e, the engineered $\Omega_{13}$ is well-fitted by Eq. (3.19). Further experiments under additional microwave bursts demonstrate that the vibrational excitations (phonons) can be coherently transferred between $R_1$ and $R_3$, with $R_2$ mediating their interactions. Using similar devices, classical Rabi oscillations and Ramsey fringes have been achieved, which are electrically tunable as well, offering a promising future for information processing based on nanomechanical vibrations[23].

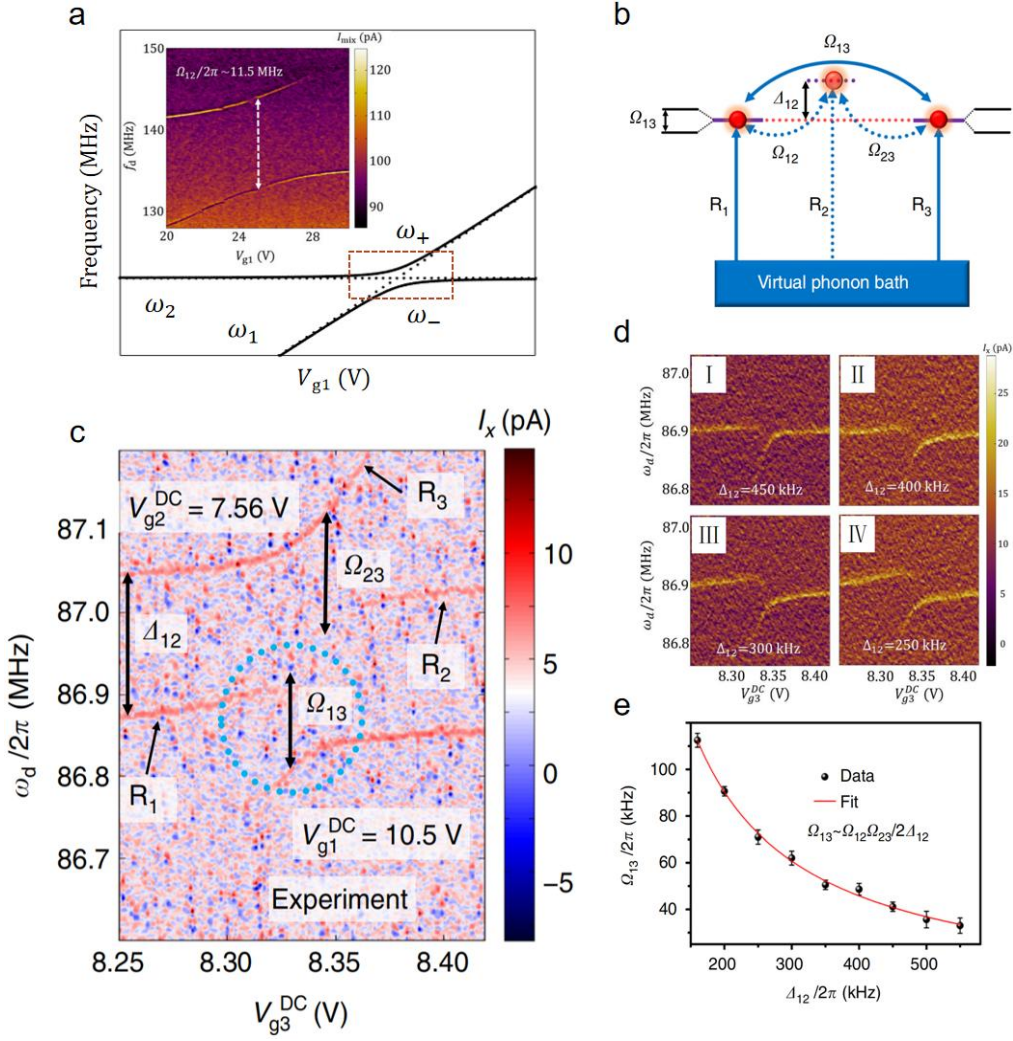

**Figure 5. Linear coupling.** (a) Resonant frequencies of two mechanical modes ($\omega_1$ and $\omega_2$) as a function of gate voltage. The two modes hybridize at the avoided crossing regime, forming two eigenmodes with frequencies of $\omega_\pm$. Inset: experimental result obtained from graphene nanomechanical resonators, corresponding to the region labeled by the red dashed box. (b) Schematic illustration of generating a tunable indirect coupling between the resonators $R_1$ and $R_3$, using $R_2$ in between as a medium. (c) Measured resonant frequencies of three coupled graphene nanomechanical resonators as a function of gate voltage. The blue dotted circle highlights that an indirect coupling $\Omega_{13}$ is generated. (d) Tuning indirect coupling strength $\Omega_{13}$ upon electrically varying detuning $\Delta_{12}$. (e) Extracted $\Omega_{13}$ as a function of $\Delta_{12}$, which is well-fitted using Eq. (3.19). (a) Reproduced under the terms of a creative commons CC-BY international license.[23] Copyright 2020, The Authors, published by National Academy of Sciences of America. (b)-(e) Reproduced under the terms of a creative commons CC-BY international license.[27] Copyright 2018, The Authors, published by Springer Nature.

Extending the last term in Eq. (3.18) to a more general form, gives birth to the nonlinear coupling[39]:

$$U_{12}^{nl} = m_0 g_{12}^{(nm)} x_1^n x_2^m, \qquad (3.20)$$

where the resonant condition is replaced by $n\omega_1 = m\omega_2$, instead of $\omega_1 = \omega_2$.

For instance, Fig. 6a shows the case of $n = 3$ and $m = 1$. The underlying physics can be easier to understand if a quantum methodology is used to transform $x_i$ to annihilation ($a_i$) and creation operators ($a_i^\dagger$) as[39]:

$$x_i = \left(\frac{\hbar}{2m_0\omega_i}\right)^{\frac{1}{2}}(a_i + a_i^\dagger). \tag{3.21}$$

Therefore, the coupling term is translated to $(a_1^\dagger)^3 a_2 + a_2^\dagger a_1^3$. The first (second) term describes the process of creating (annihilating) three mechanical excitations of mode 1 while annihilating (creating) one of mode 2 to satisfy the energy conservation $3\hbar\omega_1 = \hbar\omega_2$.

In graphene nanomechanical resonators, the nonlinear resonant condition can be reached by electrical gating or/and driving in the Duffing nonlinearity regime[21, 54, 122, 123]. Figure 6b shows experimental results of a time-resolved ring-down measurement from a graphene resonator, where the influence of nonlinear coupling is revealed ($n: m = 3: 1$). After the strong external driving is switched off at $t = 0$, the amplitude of mode 1 (with frequency of $\omega_1$) decays. Interestingly, different from the expectation of an overall exponential behavior, the energy dissipation exhibits two stages with different decay rates[21]. As illustrated in the right panel, in stage I, the vibrational amplitude is large, resulting in a strong nonlinear mode coupling $g_{\text{eff}}$ (note that the nonlinear coupling is positively related to $x_i$ in Eq. (3.20)). Therefore, it can be roughly understood that the energy dissipation channel of mode 1 has two contributions: one is directly decay to the environment with a rate of $g_1$; the other is transferred to mode 2 and then decay to the environment with a rate of $g_2$ (usually, as a higher mode, $g_2$ is larger than $g_1$). While in stage II, the decreased vibrational amplitude suppresses the nonlinear coupling, effectively isolating the two modes. As a consequence, the energy dissipation of mode 1 is only contributed by $g_1$, so that a lower decay rate is observed in stage II.

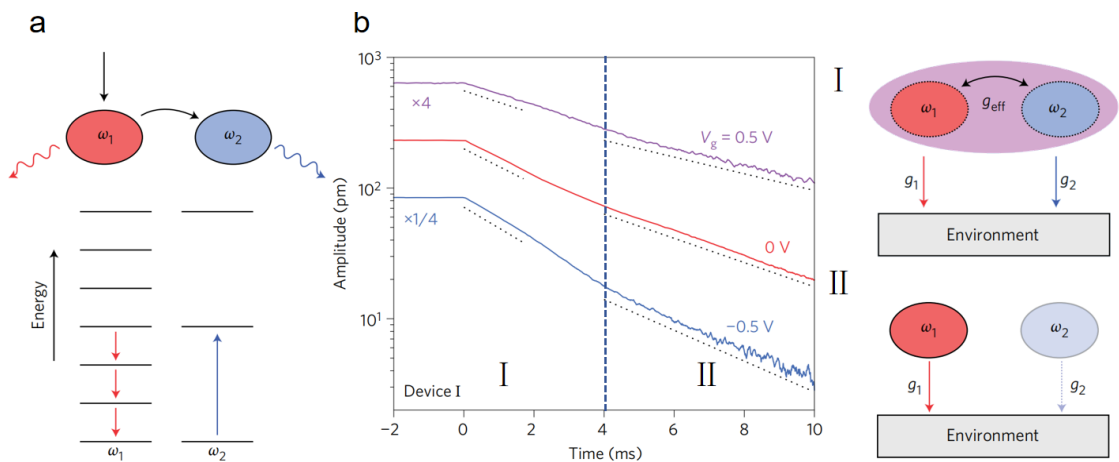

**Figure 6. Nonlinear coupling.** (a) Schematic of nonlinear coupling with $\omega_2/\omega_1 = 3$. (b) Time-resolved ring-down measurement of a graphene nanomechanical resonator, where the energy dissipation exhibits two stages with different decay rates. The corresponding schematics of stages I and II are illustrated in the right panel, respectively. (a)-(b) Reproduced with permission.[21] Copyright 2017, Springer Nature.

Another coupling mechanism is parametric coupling[52, 118], which has no requirement of $\omega_2/\omega_1$. This is realized by applying an external pump at the frequency of $\omega_p$ to the system. Although there is no restriction on $\omega_p$ in principle, it is often to choose $\omega_p = \omega_1 + \omega_2$ or $\omega_p = |\omega_1 - \omega_2|$ to achieve a strong coupling in practice. The corresponding potential energy can be written as[39]:

$$U_{12}^{pump} = m_0 g_{12}^{pump}(t) x_1 x_2. \tag{3.22}$$

Note that $g_{12}^{pump}(t)$ is time dependent, which is positively related to the magnitude of the pump signal. This parametric coupling mechanism is similar to what happens in optomechanical systems, which is used to cool or heat vibrational excitations (phonons)[124]. Graphene-based optomechanics has also been realized by coupling graphene nanomechanical resonators to superconducting microwave cavities[22, 38, 75], demonstrating optomechanically induced absorption and reflection. Here, we mainly focus on the parametric coupling between mechanical modes.

Assuming there are two mechanical vibrational modes with frequencies of $\omega_1$ and $\omega_2$ ($\omega_2 > \omega_1$), respectively. As shown in Fig. 7a, when a red-pump at $\omega_p = \omega_2 - \omega_1$ is applied, a sideband at $\omega_2 - \omega_p$ is generated and is on resonance with the mode at $\omega_1$. If the pump is sufficiently strong, avoided crossing emerges between $\omega_1$ and $\omega_2 - \omega_p$, similar to the case of linear coupling (left panel of Fig. 7b). The corresponding mode splitting $2g$ is linearly scaled with the pump voltage (right panel of Fig. 7b)[52]. Alternatively, when a blue-pump at $\omega_p = \omega_1 + \omega_2$ is applied, it amplifies both modes (Fig. 7c). As shown in Fig. 7d, when increasing the pump amplitude, the resonance peak of mode 1 (with the frequency of $\omega_1$) becomes narrower (left panel), with the effective dissipation suppressed (right panel)[52]. This result under the blue-pump can be viewed as a generalized parametric amplification of a single mode at $\omega$ with a $2\omega$ pump[47, 52, 125, 126].

In graphene nanomechanical resonators, the exceptional frequency tunability enables a further three-mode alignment (with frequencies from low to high as $\omega_1$, $\omega_2$ and $\omega_3$ having $\omega_3 = \omega_1 + \omega_2$)[118]. As shown in Fig. 7e, using the language of optomechanics, the highest mode at $\omega_3$ serves as the cavity mode ($\omega_c = \omega_3$). The red-pump is applied at the sideband of the cavity having $\omega_p = \omega_{sb} = \omega_c - \omega_1$. The three-mode alignment makes $\omega_p$ overlap with $\omega_2$ so that the pumping is enhanced by the mechanical quality factor $Q_2$ of the second mode. Using this strategy, sideband cooling the thermal motion of the lowest mode is demonstrated (Fig. 7f)[118].

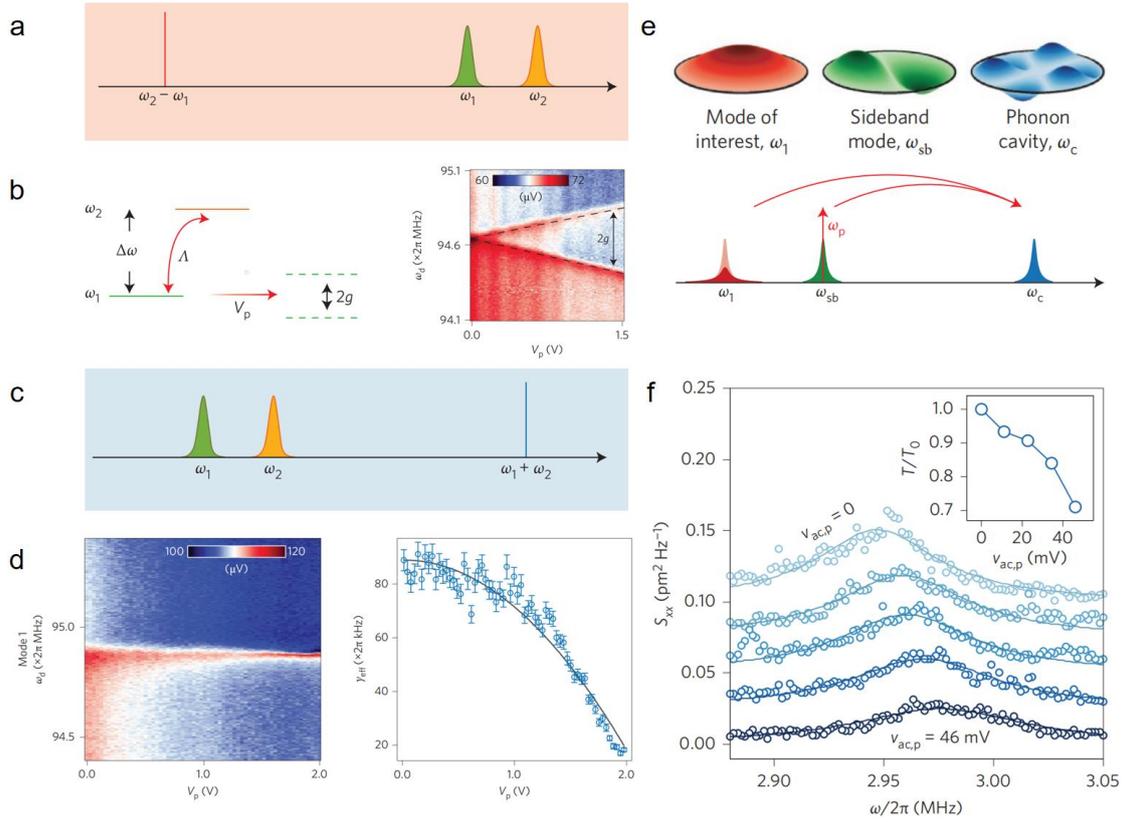

**Figure 7. Parametric coupling.** (a) Schematic of red-pump at $\omega_p = \omega_2 - \omega_1$, with $\omega_1$ and $\omega_2$ as the resonant frequencies of two mechanical modes, respectively, having $\omega_2 > \omega_1$. (b) A sideband at $\omega_2 - \omega_p$ is generated and is on resonance with the mode at $\omega_1$, resulting in a mode splitting $2g$ linearly scaled with the pump voltage $V_p$. (c) Schematic of blue-pump at $\omega_p = \omega_1 + \omega_2$. (d) The resonance peak (the mode at $\omega_1$) becomes narrower when increasing the blue-pump amplitude, indicating the amplification of the mode. (e) Schematic of red-pump for the case of three-mode alignment. (f) Measured spectral noise density of the lowest mode (with the frequency of $\omega_1$) at different red-pump amplitudes, demonstrating sideband cooling of the thermal mechanical motion. (a)-(d) Reproduced with permission.[52] Copyright 2016, Springer Nature. (e)-(f) Reproduced with permission.[118] Copyright 2016, Springer Nature.

## 4. Applications

The exceptional mechanical and electrical properties make graphene naturally an excellent candidate for hosting NEMS applications. It has been demonstrated that graphene nanomechanical resonators can serve as electro-mechanical switches[28, 29, 127-132], tunable oscillators[26], and acoustic microphones[30, 31, 133, 134]. In these applications, the 2D nature of graphene generally results in lower power consumptions and higher transduction efficiencies[41], compared with MEMS, offering a promising future for graphene-based NEMS applications.

Another important and widely-studied application is employing graphene nanomechanical resonators as various sensors of, for example, pressure[67, 119, 135-140], electromagnetic radiation[64], mass[46, 141], vibration/sound[33, 142], acceleration[34, 143-148], and force[24]. As mentioned above, the low mass density and relatively small spring constant of graphene nanomechanical resonators are favorable to generate a strong mechanical response (for example, a pronounced change in resonant frequency or static displacement) upon external excitations, which is beneficial for sensitive sensing. Here, we briefly introduce several typical examples and highlight their underlying sensing mechanisms and strategies. More in-depth discussions and comparisons of the graphene-based sensors can be found in previous reviews and references therein[43, 44].

Figure 8a shows the schematic of a graphene pressure sensor. The suspended graphene membrane defines a gas cavity that is connected with the outer environment through a venting channel. The device acts as a squeeze-film pressure sensor[119]. The pressure change effectively modulates the tension in the graphene membrane, thus leading to the resonant frequency change. Figure 8b shows the measured resonant frequency as a function of pressure, which can be well-understood based on the ideal gas law (black dotted curves)[119]. Other designs such as using sealed gas cavities without venting channels have also been demonstrated[32, 58, 137, 138]. Moreover, in addition to detecting the resonant frequency[32, 58, 67, 119, 149], which has the advantages of high resolution and optical-friendly accessibility[140], the pressure can be sensed based on the displacement-induced capacitance change[137, 138, 150, 151] or via piezoresistive effect-induced resistance change as well[135, 136, 152-154]. These strategies are employed in other sensing applications in similar ways, as discussed below.

The tension-induced resonant frequency change is also used in sensing other physical quantities. For example, Fig. 8c shows the schematic of a graphene bolometer, where the absorbed electromagnetic radiation heats the suspended graphene membrane, thereby shifting the resonant frequency via tension modulation (Fig. 8d)[64]. Besides, the resonant frequency can also be shifted if the effective mass of the resonator is varied. The low mass density of graphene thus offers high sensitivity in mass-sensing adsorbates deposited on the suspended graphene membrane[46].

Sensing via the displacement change is also straightforward. An example is illustrated in Fig. 8e, where the device is aiming to detecting the ultrasound introduced through the substrate[33]. As discussed, the nonlinear effect is sensitive to the displacement of the resonator under strong driving. Therefore, the nonlinearity-induced frequency shift is used to extract the displacement change under different ultrasound drive amplitudes, from which the influence of the ultrasound is identified (Fig. 8f)[33]. Another interesting example is shown in Fig. 8g. The motion of a single live bacterium is probed and analyzed by detecting the displacement of a graphene drum resonator on which the bacterial cell is adhered (Fig. 8h)[155].

The piezoresistive effect transforms the strain modulations into resistance changes that can be used in sensing, for example, acceleration forces[34, 143]. Figure 8i shows the schematic of an accelerometer, in which a silicon-proof mass is attached to a suspended graphene ribbon to enhance the response to acceleration forces. As shown in Fig. 8j, the resistance change is proportional to the actual acceleration[34].

In addition to the above-mentioned strategies, a recent study demonstrates the idea of using different frequency responses upon electrical gating to quantify the interfacial friction between graphene and its supporting substrate[25]. Figure 8k illustrates the schematic of three doubly clamped graphene resonators coupled in series. When cycling the gate voltage applied to the middle resonator, a reversible sliding of the clamping points is achieved. This sliding behavior feeds extra length to the middle trench that competes with the commonly observed tension increase under stretching. Experimentally, a frequency loop is observed upon gating (Fig. 8l), demonstrating that the resonant frequency is strongly influenced not only by the value of the gate voltage but also by the way how it is applied. Moreover, it is found that the loop is a measure of the nanoscale friction since its area is proportional to the energy dissipated as the membrane slides back and forth on its support[25].

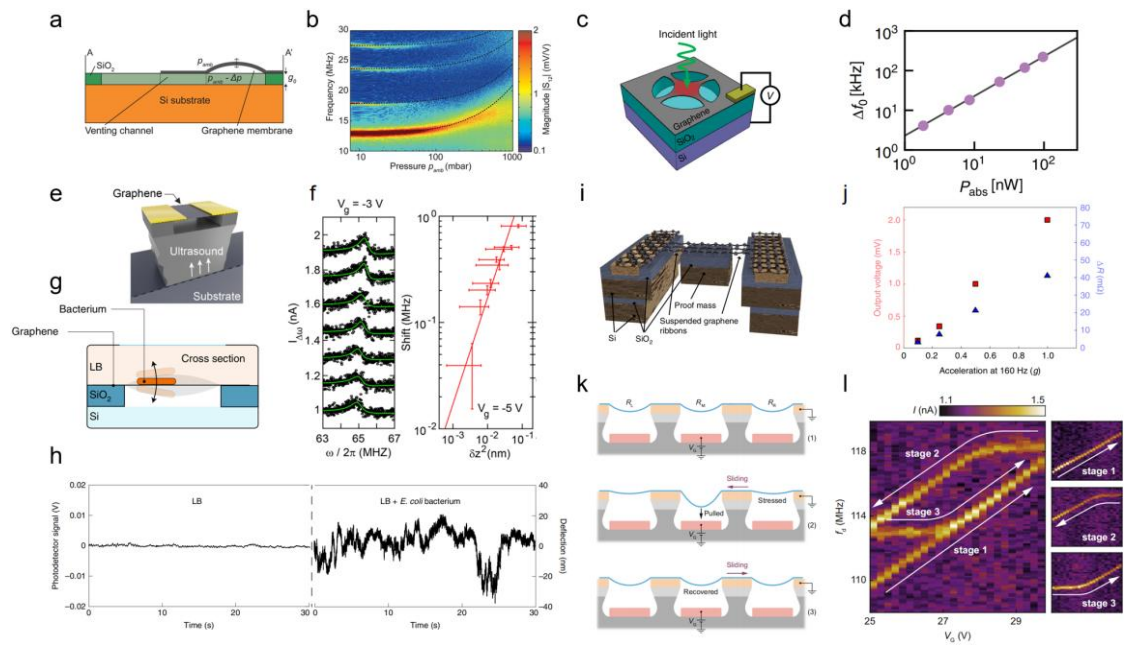

**Figure 8. Various sensors based on graphene nanomechanical resonators and their sensing strategies.** (a) Schematic of a pressure sensor. (b) The resonant frequency changes with the gas pressure. (c) Schematic of a bolometer. (d) The resonant frequency changes with the absorbed power. (e) Schematic of an ultrasound detector. (f) Different ultrasound drive amplitudes result in different membrane displacements, thereby changing the lineshape of the resonance peak in the nonlinear regime. (g) Schematic of probing a single bacterium using a graphene nanomechanical resonator. (h) The motion of a live bacterium generates larger displacement variations of the resonator to be detected. (i) Schematic of an accelerometer. (j) The piezoresistance changes with the acceleration. (k) Schematic of coupled graphene nanomechanical resonators for quantifying interfacial sliding. (l) The resonant frequency of the middle resonator exhibits a loop upon cycling the applied gate voltage, with its area proportional to the energy dissipated due to the nanoscale friction. (a)-(b) Reproduced with permission.[119] Copyright 2016, American Chemical Society. (c)-(d) Reproduced under the terms of a creative commons CC-BY international license.[64] Copyright 2019, The Authors, published by Springer Nature. (e)-(f) Reproduced with permission.[33] Copyright 2018, American Chemical Society. (g)-(h) Reproduced with permission.[155] Copyright 2022, Springer Nature. (i)-(j) Reproduced with permission.[34] Copyright 2019, Springer Nature. (k)-(l) Reproduced under the terms of a creative commons CC-BY international license.[25] Copyright 2022, The Authors, published by Springer Nature.

Arranging coupled mechanical resonators into a large-scale periodic array, a vibrational band emerges, giving rise to the concept of phononic crystal[156]. Engineering such a periodic architecture leads to a variety of promising mechanical applications, including realizing mechanical waveguides[157-159], manipulating mechanical dissipations[160-165], and generating topological acoustic/mechanical systems[166-170]. Employing graphene as the material platform for hosting phononic crystals offers the powerful capability of *in-situ* electrical tuning of the phononic band structure after the device is fabricated. This capability not only enables realizing on-chip phononic phase transitions, but also allows achieving tunable acoustic shields to electrically engineer mechanical dissipations[165].

Two different experimental approaches have been explored towards graphene-based tunable phononic crystals. As shown in Fig. 9a, in the first approach[35, 171, 172], the suspended graphene membrane is tailored with a honeycomb lattice of holes using a focus ion beam. This design leads to the formation of a phononic band with a band gap at ~30 MHz (Fig. 9b). The optically measured result is shown in Fig. 9c, demonstrating that the phononic band can be tuned by the global back gate[171]. As shown in Fig. 9d, in the second approach, the suspended graphene membrane is clamped by many prepatterned pillars forming a regular array[36]. Since it is more technically friendly to carve a periodic pattern in the rigid substrate than in a suspended membrane, this design allows a smaller lattice constant and a local back gate for electrical tuning. The measured frequency spectrum demonstrates a quasi-continuous band-like feature approaching GHz (from ~120 MHz to ~980 MHz)[36], which is one order of magnitude higher. The local back gate not only provides a strong tunability of the band frequency but also enables the electrical readout using a frequency mixing technique (Fig. 9e)[36]. A recent study further proposes to take the advantage of fabrication flexibility in this design to realize a triangular lattice of the standing pillars, so that the topological acoustics/mechanics can be brought to the on-chip scale (Fig. 9f)[173].

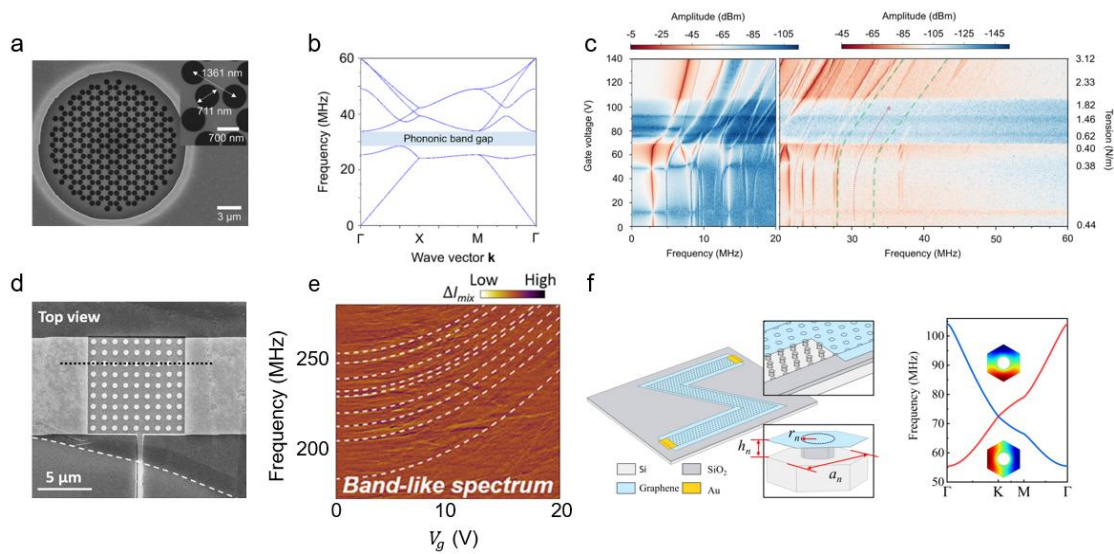

**Figure 9. Towards graphene-based tunable phononic crystals.** (a) Scanning electron microscope (SEM) image of a graphene phononic crystal device, realized by selectively patterning the membrane with a periodic array of holes. (b) Calculated phononic band. (c) Measured frequency

spectrum, showing the electrically tunable vibrational phononic band. (d) SEM image of a graphene phononic crystal device, realized by pinning the membrane to an array of pre-patterned standing pillars. (e) Measured band-like frequency spectrum, which can be electrically tuned by the gate voltage. (f) Schematic of an on-chip topological acoustic device based on graphene (left panel) and its band diagram (right panel). (a)-(c) Reproduced under the terms of a creative commons CC-BY international license.[171] Copyright 2023, The Authors, published by Institute of Physics Publishing. (d)-(e) Reproduced with permission.[36] Copyright 2021, American Chemical Society. (f) Reproduced with permission.[173] Copyright 2024, American Physical Society.

Moreover, graphene, as an atomically thin material, significantly down-scales the characteristic size of nanomechanical devices it hosts, compared with MEMS devices. This leads to high resonant frequencies while maintaining sufficiently high quality factors at low temperature[22-24], which is beneficial for a lower vibrational phonon occupation approaching the quantum limit. Meanwhile, it is expected that the zero-point motion, which reflects the quantum nature of the resonator, is large in graphene nanomechanical devices[38, 75], due to the low spring constant and mass density. Besides, the large mechanical response to excitations makes graphene nanomechanical resonators favorable to serve as interfaces to various physical systems to transduce quantum information. Therefore, it is interesting and promising to explore potential quantum applications of graphene nanomechanical resonators.

Here, we review several representative examples towards this goal. The first one is well-studied optomechanical systems which couple graphene nanomechanical resonators to superconducting microwave cavities[21, 22, 24, 38, 75, 86, 87]. The left panel of Fig. 10a shows a typical device and its cross-sectional schematic. Through sideband cooling, the phonon occupation is cooled to $7.2 \pm 0.2$ at 15 mK (right panel)[24], approaching the quantum ground state of 1. In addition, force sensitivity of $390 \pm 30$ zN Hz$^{-1/2}$ is achieved[24], holding a promising future for quantum sensing of electron and nuclear spins of molecules adsorbed on the surface of graphene[174]. The left panel of Fig. 10b shows a different type of optomechanical device that a graphene membrane is suspended above nitrogen-vacancy (NV) centers, which are stable single-photon emitters embedded in nanodiamonds. The emission of the NV center is strongly modulated upon electrically pulling the membrane due to active control of their optomechanical coupling strength (right panel in Fig. 10b)[175].

The excellent electrical property of graphene also allows electromechanical coupling between the mechanical vibration and quantum electronic states[18, 37]. To obtain the electrostatic energy in Eq. (3.8), a classical picture of a metallic capacitor with a large density of states is assumed. As a consequence, the electrochemical potential $\mu$ is considered to be constant. However, when the device is operated in a low-density regime, the influence of changes in $\mu$ plays a pronounced role. For example, Fig. 10c shows a graphene nanomechanical resonator under a large magnetic field, where discrete Landau levels are formed. In such a regime, the capacitive force is given as[18]:

$$F = -\frac{1}{2}\left(V_\text{g} - \frac{\mu}{e}\right)^2 \left(1 - 2\frac{C_\text{g}}{C_\text{Q}}\right)\frac{\text{d}C_\text{g}}{\text{d}z}, \quad (4.1)$$

where $C_Q = Ae^2 \, dn/d\mu$ is the quantum capacitance, with $A$ is the sample size, $e$ is the electron charge, and $n$ is the charge density. Note that when $\mu$ is fixed to be zero and thus $d\mu/dn = 0$, Eq. (4.1) is simplified to $F = -\frac{1}{2}\frac{dC_g}{dz}V_g^2$, which is the case of metallic capacitance assumption. This modification results in additional resonant frequency shifts. As shown in the right panel in Fig. 10c, with proper device parameters, the frequency shifts are dominated by the shift in $\mu$ (with less pronounced influences from terms related to $dn/d\mu$) induced by the applied magnetic field, enabling direct access to electrochemical potentials of electrons[18]. This provides the possibility of nanomechanical readout of exotic quantum states, such as fractional quantum Hall effects[176, 177], and Wigner crystallization[178]. Another similar example is demonstrated in Fig. 10d. In this device, the suspended graphene nanoribbon is ~50 nm in width[48]. The transport behavior is thus similar to that in a confined 1D quantum dot system[78, 179-183], where single electron tunneling is dominating. As shown in the right panel in Fig. 10d, when an electron tunnels through the quantum dot, the resonant frequency develops a dip. The underlying physical mechanism is quite similar to the previous one: the change in $\mu$ modulates the average electron occupation in the dot[48] (the electron density in the previous case[18]) and thus the electrostatic force, thereby shifting the resonant frequency. Here, the shift in $\mu$ contributes to the overall frequency-voltage tuning, while $dn/d\mu$ terms produce profound frequency dips.

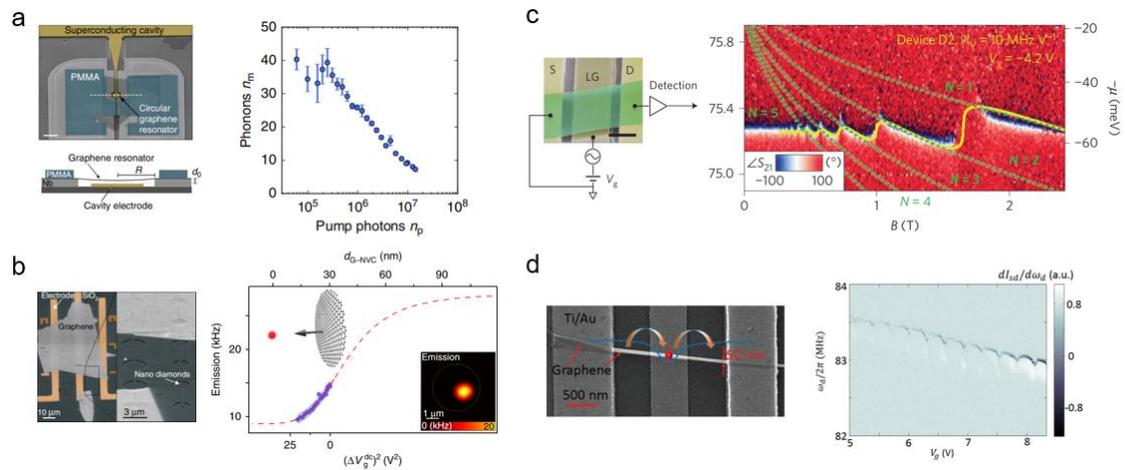

**Figure 10. Towards quantum applications based on graphene nanomechanical resonators.** (a) Coupling a graphene nanomechanical resonator to a superconducting microwave cavity for optomechanical sideband cooling. (b) Suspending a graphene membrane over NV centers for electromechanical control of the photon emission. (c) Operating a graphene nanomechanical resonator in the presence of Landau level formation, where the modulation of electrochemical potentials shifts the resonant frequency. (d) Coupling the mechanical vibration of a suspended graphene nanoribbon to single electron tunneling through the device. (a) Reproduced under the terms of a creative commons CC-BY international license.[24] Copyright 2016, The Authors, published by Springer Nature. (b) Reproduced under the terms of a creative commons CC-BY international license.[175] Copyright 2016, The Authors, published by Springer Nature. (c) Reproduced with permission.[18] Copyright 2016, Springer Nature. (d) Reproduced with permission.[48] Copyright 2017, Royal Society of Chemistry.

## 5. Future prospects

While graphene nanomechanical resonators have demonstrated great potential in serving as next-generation NEMS, and for tunable classical/quantum phononic device applications, there are still challenges, as well as opportunities, lying ahead.

The most important challenge is, probably, increasing the reproducibility of devices. This requirement has two-fold meanings. The first one is to achieve a precise control of, for example, strain distribution in transferring graphene flakes, to realize high uniformity from device to device. This is not trivial since most of the current 2D material transfer techniques[184-188] are developed for 2D electronic and/or optical devices. In these devices, the most care is taken to avoid bubbles and residues at the interfaces[72, 187-192], while the influence of the strain distributed in the 2D flakes receives less attention[193, 194]. However, in nanomechanical resonators, non-uniformly distributed strain plays a significant role in determining the vibrational modes[50, 55, 59, 195], thus their frequencies, nonlinearities, and couplings with other modes. (A particular example is the formation of wrinkles, which appears quite often when transferring 2D materials, especially as a suspended structure[74, 196-201]. However, a clear physical picture of these nanoscale wrinkles, including their creation, dynamics, and annihilation, is still missing.) Recently, similar attentions have been driven to strain variations in stacking twisted 2D moiré materials due to their crucial role in determining the uniformity of the moiré superlattice[202-205]. With these requirements, more advanced 2D material transfer techniques are invited, compared with state-of-the-art (still roughly rely on macroscale control on the transfer process) techniques. We would like to point out that a high reproducibility from device to device is essential for a deeper understanding of mechanical dissipation in graphene nanomechanical resonators, since systematic studies with well-controlled experiments are expected. The second fold is to achieve a high reproducibility so that large-scale integration of devices can be within reach. This not only requires control in the transfer procedure but also in the scalable synthesis of graphene with uniform quality[206-209], thus being more challenging.

In addition to challenges, we would like to list several future opportunities. As a 2D van der Waals material, it is convenient to integrate graphene with other materials to form heterostructures without considering lattice mismatch[15, 210]. Nanomechanical devices made of these graphene-based heterostructures[97, 211-223] offer opportunities for various applications.

Taking integration with other 2D materials as examples first, Fig. 11a shows nanomechanical resonators based on graphene-$MoS_2$ heterostructures, whose frequency-voltage tuning spectrum exhibits distinct kinks and emerges hysteresis depending on the sweep direction of the gate voltage[213]. A similar result is demonstrated in resonators based on graphene-hBN heterostructures. As shown in Fig. 11b, frequency kinks are also observed, and the electrical gating effect on the resonant frequency clearly deviates from the prediction of a continuum mechanics model[215]. The underlying physical origin is not fully understood. Possible candidates[60, 97, 213, 215] include interlayer sliding, and the influence of bubbles and folds at the interfaces. It is worth noting that even in nanomechanical resonators based on bilayer and few-layer graphene, frequency jumps are observed upon gating[98]. This result is understood as the creation and annihilation of solitons, which are local changes in domain walls[98, 202], suggesting a pronounced interfacial influence on mechanics. More recently, in nanomechanical resonators made of twisted bilayer graphene, a controllable hysteretic response is reported, which is attributed to the viscoelasticity originating from interfacial sliding between two incommensurate carbon layers[224]. These observations, along with the recent demonstration of sliding nanomechanical resonators[25], indicate that graphene-based nanomechanical resonators can serve as an ideal platform to access unconventional mechanics at their atomically-smooth interfaces.

Figure 11c shows another example that employing nanomechanical resonators made of graphene-based 2D heterostructures to investigate phase transitions. The heterostructure consists of monolayer $WSe_2$, bilayer $CrI_3$, and few-layer graphene. Here, the graphene layer plays two roles: protecting the air-sensitive 2D material from degradation[211, 217] and serving as a conducting electrode for electrical actuation[217]. The resonant frequency of the device exhibits hysteretic jumps when varying the external magnetic field to trigger an antiferromagnetic to ferromagnetic phase transition in bilayer $CrI_3$. This is because the magnetostriction effect couples the magnetic and mechanical degrees of freedom, generating a strain change, thus a frequency jump when the phase transition is experienced[217]. Moreover, the strain-tuning of the magnetic interaction is demonstrated through the inverse magnetostriction effect[217].

In addition to 2D materials, graphene can be integrated with conventional bulk materials for nanomechanical applications as well. For example, a graphene membrane can be transferred onto a larger suspended silicon nitride ($SiN_x$) membrane with predefined holes[212, 214, 216, 218, 222]. The two mechanical systems are orders of magnitude different in mass, while are electrically tuned on resonance with similar frequencies. Although their vibrational amplitudes differ in orders of magnitude, a strong mode coupling is achieved[212], resulting in a giant nonlinearity generated in $SiN_x$ resonators[218]. The strong nonlinearity is also used to demonstrate phononic frequency combs upon parametrically driving the hybridized modes (Fig. 11d)[218]. Figure 11e shows a different scenario in which a graphene membrane is transferred and clamped onto a micromachined silicon comb-drive actuator. This device can be used to controllably engineer the strain field in the graphene membrane[221, 225]. Alternatively, the graphene membrane serves as a mechanical element in mediating a tunable mechanical coupling between the comb-drive actuator and a suspending silicon beam[226].

Finally, an interesting direction towards quantum applications is to realize nanomechanical qubits[227-229]. This can be done by strongly coupling nanomechanical resonators to quantum dots embedded[227]. Although suspended graphene quantum dots have been demonstrated[48, 230, 231], the device performance still needs improvements. Recently, significant efforts have been devoted to unsuspended graphene quantum dot devices[232], in which excellent electrical tunability[233-238] as well as the capability of manipulating various electronic quantum degrees of freedom have been achieved[239-243]. This offers a promising future for graphene-based nanomechanical qubits, provided by high resonant frequency[18-20], low mechanical loss[22-24], large zero-point motion[38, 75], and excellent coherency of vibrational phonons[23] in graphene nanomechanical resonators at low temperature. Moreover, in addition to the charge degree of freedom[244], additional electronic degrees of freedom (e.g., spin[245-250], valley[251-254], minivalley[255, 256]) in graphene give the possibility of generating novel coupling schemes[257, 258] with mechanical degree of freedom in suspended quantum dot devices.

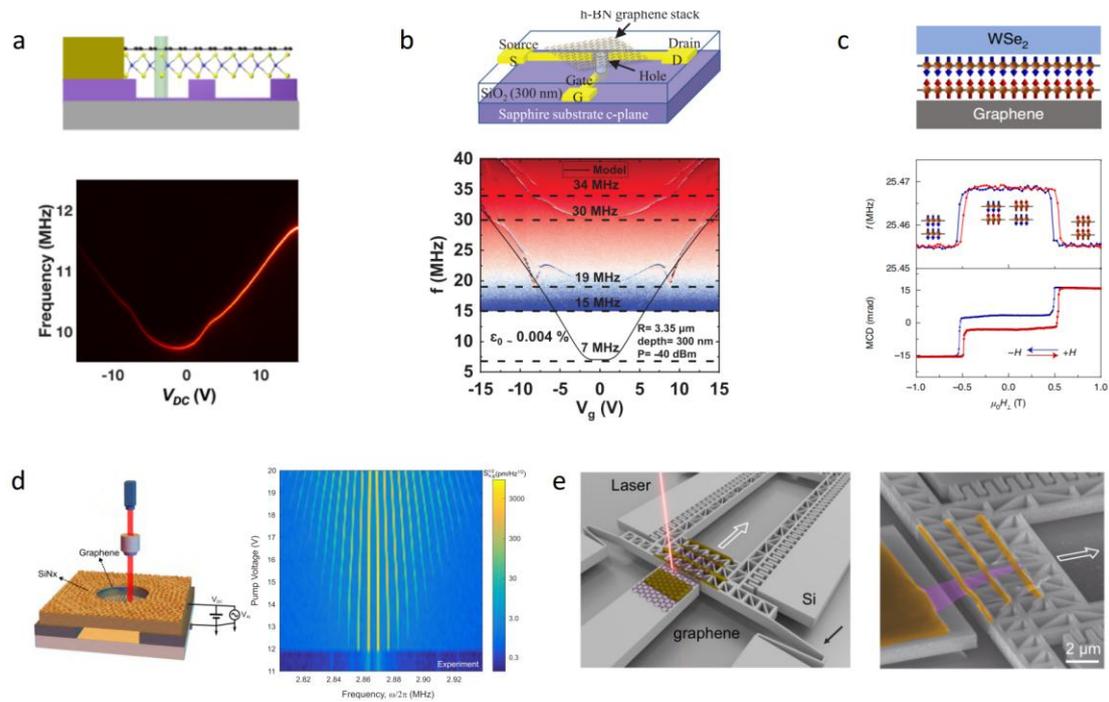

**Figure 11. Nanomechanical devices made of graphene-based heterostructures.** (a) A nanomechanical resonator based on graphene-$MoS_2$ heterostructure, whose frequency-voltage tuning spectrum exhibits distinct kinks. (b) A nanomechanical resonator based on graphene-hBN heterostructure, in which the electrical gating effect on the resonant frequency deviates from the prediction of a continuum mechanics model. (c) A nanomechanical resonator made of the heterostructure consisting of monolayer $WSe_2$, bilayer $CrI_3$, and few-layer graphene, which is used for studying an antiferromagnetic to ferromagnetic phase transition in bilayer $CrI_3$. (d) A mechanical resonator based on graphene-$SiN_x$ heterostructure, demonstrating a phononic frequency comb upon parametrical driving. (e) A micromachined silicon comb-drive actuator with an integrated clamped graphene membrane. (a) Reproduced with permission.[213] Copyright 2018, American Chemical Society. (b) Reproduced with permission.[215] Copyright 2020, AIP Publishing. (c) Reproduced with permission.[217] Copyright 2020, Springer Nature. (d) Reproduced with permission.[218] Copyright 2020, American Chemical Society. (e) Reproduced with permission.[221] Copyright 2018, American Chemical Society.

To conclude, after more than 15 years of studies since the first demonstration of graphene nanomechanical resonators[17], this field is far from being extensively explored. There are still plenty of opportunities in deepening fundamental understanding of mesoscopic dynamics of nanomechanical vibrations, engineering figure-of-merit of nanomechanical devices, and developing diverse nanomechanical/phononic applications. With more efforts being devoted, a more promising future of tunable graphene nanomechanical resonators can be expected.

**Acknowledgements**

This work was supported by the Natural Science Foundation of Jiangsu Province (Grant No. BK20240123), the National Key Research and Development Program of China (Grant No. 2022YFA1405900), and the National Natural Science Foundation of China (Grant Nos. 12274397, 12274401, and 12034018).

**References**

[1] Novoselov K S, Geim A K, Morozov S V, Jiang D, Zhang Y, Dubonos S V, Grigorieva I V and Firsov A A 2004 *Science* **306** 666

[2] Geim A K and Novoselov K S 2007 *Nat. Mater.* **6** 183

[3] Novoselov K S, Geim A K, Morozov S V, Jiang D, Katsnelson M I, Grigorieva I V, Dubonos S V and Firsov A A 2005 *Nature* **438** 197

[4] Zhang Y, Tan Y-W, Stormer H L and Kim P 2005 *Nature* **438** 201

[5] Bolotin K I, Sikes K J, Jiang Z, Klima M, Fudenberg G, Hone J, Kim P and Stormer H L 2008 *Solid State Commun.* **146** 351

[6] Du X, Skachko I, Barker A and Andrei E Y 2008 *Nat. Nanotechnol.* **3** 491

[7] Nair R R, Blake P, Grigorenko A N, Novoselov K S, Booth T J, Stauber T, Peres N M R and Geim A K 2008 *Science* **320** 1308

[8] Wang F, Zhang Y, Tian C, Girit C, Zettl A, Crommie M and Shen Y R 2008 *Science* **320** 206

[9] Bonaccorso F, Sun Z, Hasan T and Ferrari A C 2010 *Nat. Photonics* **4** 611

[10] Lee C, Wei X, Kysar J W and Hone J 2008 *Science* **321** 385

[11] Blees M K, Barnard A W, Rose P A, Roberts S P, McGill K L, Huang P Y, Ruyack A R, Kevek J W, Kobrin B, Muller D A and McEuen P L 2015 *Nature* **524** 204

[12] Wei Y and Yang R 2019 *Natl. Sci. Rev.* **6** 324

[13] Castro Neto A H, Guinea F, Peres N M R, Novoselov K S and Geim A K 2009 *Rev. Mod. Phys.* **81** 109

[14] Lee G-H, Cooper R C, An S J, Lee S, van der Zande A, Petrone N, Hammerberg A G, Lee C, Crawford B, Oliver W, Kysar J W and Hone J 2013 *Science* **340** 1073

[15] Akinwande D, Huyghebaert C, Wang C-H, Serna M I, Goossens S, Li L-J, Wong H S P and Koppens F H L 2019 *Nature* **573** 507

[16] Romagnoli M, Sorianello V, Midrio M, Koppens F H L, Huyghebaert C, Neumaier D, Galli P, Templ W, D'Errico A and Ferrari A C 2018 *Nat. Rev. Mater.* **3** 392

[17] Bunch J S, van der Zande A M, Verbridge S S, Frank I W, Tanenbaum D M, Parpia J M, Craighead H G and McEuen P L 2007 *Science* **315** 490

[18] Chen C, Deshpande V V, Koshino M, Lee S, Gondarenko A, MacDonald A H, Kim P and Hone J 2016 *Nat. Phys.* **12** 240

[19] Eichler A, Moser J, Chaste J, Zdrojek M, Wilson-Rae I and Bachtold A 2011 *Nat. Nanotechnol.* **6** 339

[20] Jung M, Rickhaus P, Zihlmann S, Eichler A, Makk P and Schönenberger C 2019 *Nanoscale* **11** 4355

[21] Güttinger J, Noury A, Weber P, Eriksson A M, Lagoin C, Moser J, Eichler C, Wallraff A, Isacsson A and Bachtold A 2017 *Nat. Nanotechnol.* **12** 631

[22] Weber P, Güttinger J, Tsioutsios I, Chang D E and Bachtold A 2014 *Nano Lett.* **14** 2854

[23] Zhang Z Z, Song X X, Luo G, Su Z J, Wang K L, Cao G, Li H O, Xiao M, Guo G C, Tian L, Deng G W and Guo G P 2020 *Proc. Natl. Acad. Sci.* **117** 5582

[24] Weber P, Güttinger J, Noury A, Vergara-Cruz J and Bachtold A 2016 *Nat. Commun.* **7** 12496

[25] Ying Y, Zhang Z-Z, Moser J, Su Z-J, Song X-X and Guo G-P 2022 *Nat. Commun.* **13** 6392

[26] Chen C, Lee S, Deshpande V V, Lee G-H, Lekas M, Shepard K and Hone J 2013 *Nat. Nanotechnol.* **8** 923

[27] Luo G, Zhang Z-Z, Deng G-W, Li H-O, Cao G, Xiao M, Guo G-C, Tian L and Guo G-P 2018 *Nat.*

*Commun.* **9** 383

[28] Liu X, Suk J W, Boddeti N G, Cantley L, Wang L, Gray J M, Hall H J, Bright V M, Rogers C T, Dunn M L, Ruoff R S and Bunch J S 2014 *Adv. Mater.* **26** 1571

[29] Milaninia K M, Baldo M A, Reina A and Kong J 2009 *Appl. Phys. Lett.* **95** 183105

[30] Zhou Q and Zettl A 2013 *Appl. Phys. Lett.* **102** 223109

[31] Abrahams M P, Martinez J, Steeneken P G and Verbiest G J 2024 *Nano Lett.* **24** 14162

[32] Lee M, Davidovikj D, Sajadi B, Šiškins M, Alijani F, van der Zant H S J and Steeneken P G 2019 *Nano Lett.* **19** 5313

[33] Verbiest G J, Kirchhof J N, Sonntag J, Goldsche M, Khodkov T and Stampfer C 2018 *Nano Lett.* **18** 5132

[34] Fan X, Forsberg F, Smith A D, Schröder S, Wagner S, Rödjegård H, Fischer A C, Östling M, Lemme M C and Niklaus F 2019 *Nat. Electron.* **2** 394

[35] Kirchhof J N, Weinel K, Heeg S, Deinhart V, Kovalchuk S, Höflich K and Bolotin K I 2021 *Nano Lett.* **21** 2174

[36] Zhang Q-H, Ying Y, Zhang Z-Z, Su Z-J, Ma H, Qin G-Q, Song X-X and Guo G-P 2021 *Nano Lett.* **21** 8571

[37] Singh V, Irfan B, Subramanian G, Solanki H S, Sengupta S, Dubey S, Kumar A, Ramakrishnan S and Deshmukh M M 2012 *Appl. Phys. Lett.* **100** 233103

[38] Singh V, Bosman S J, Schneider B H, Blanter Y M, Castellanos-Gomez A and Steele G A 2014 *Nat. Nanotechnol.* **9** 820

[39] Bachtold A, Moser J and Dykman M I 2022 *Rev. Mod. Phys.* **94** 045005

[40] Xu B, Zhang P, Zhu J, Liu Z, Eichler A, Zheng X-Q, Lee J, Dash A, More S, Wu S, Wang Y, Jia H, Naik A, Bachtold A, Yang R, Feng P X L and Wang Z 2022 *ACS Nano* **16** 15545

[41] Ferrari P F, Kim S and van der Zande A M 2023 *Appl. Phys. Rev.* **10** 031302

[42] Steeneken P G, Dolleman R J, Davidovikj D, Alijani F and van der Zant H S J 2021 *2D Mater.* **8** 042001

[43] Lemme M C, Wagner S, Lee K, Fan X, Verbiest G J, Wittmann S, Lukas S, Dolleman R J, Niklaus F, van der Zant H S J, Duesberg G S and Steeneken P G *Research* 2020 8748602

[44] Fan X, He C, Ding J, Gao Q, Ma H, Lemme M C and Zhang W 2024 *Microsyst. Nanoeng.* **10** 154

[45] Castellanos-Gomez A, Singh V, van der Zant H S J and Steele G A 2015 *Ann. Phys.* **527** 27

[46] Chen C, Rosenblatt S, Bolotin K I, Kalb W, Kim P, Kymissis I, Stormer H L, Heinz T F and Hone J 2009 *Nat. Nanotechnol.* **4** 861

[47] Su Z-J, Ying Y, Song X-X, Zhang Z-Z, Zhang Q-H, Cao G, Li H-O, Guo G-C and Guo G-P 2021 *Nanotechnology* **32** 155203

[48] Luo G, Zhang Z-Z, Deng G-W, Li H-O, Cao G, Xiao M, Guo G-C and Guo G-P 2017 *Nanoscale* **9** 5608

[49] Xu Y, Chen C, Deshpande V V, DiRenno F A, Gondarenko A, Heinz D B, Liu S, Kim P and Hone J 2010 *Appl. Phys. Lett.* **97** 243111

[50] Garcia-Sanchez D, van der Zande A M, Paulo A S, Lassagne B, McEuen P L and Bachtold A 2008 *Nano Lett.* **8** 1399

[51] Guan F, Kumaravadivel P, Averin D V and Du X 2015 *Appl. Phys. Lett.* **107** 193102

[52] Mathew J P, Patel R N, Borah A, Vijay R and Deshmukh M M 2016 *Nat. Nanotechnol.* **11** 747

[53] Davidovikj D, Alijani F, Cartamil-Bueno S J, van der Zant H S J, Amabili M and Steeneken P G 2017 *Nat. Commun.* **8** 1253

[54] Keşkekler A, Shoshani O, Lee M, van der Zant H S J, Steeneken P G and Alijani F 2021 *Nat. Commun.* **12** 1099

[55] Barton R A, Ilic B, van der Zande A M, Whitney W S, McEuen P L, Parpia J M and Craighead H G 2011 *Nano Lett.* **11** 1232

[56] Davidovikj D, Poot M, Cartamil-Bueno S J, van der Zant H S J and Steeneken P G 2018 *Nano Lett.* **18** 2852

[57] Ye F, Lee J and Feng P X L 2018 *Nano Lett.* **18** 1678

[58] Bunch J S, Verbridge S S, Alden J S, van der Zande A M, Parpia J M, Craighead H G and McEuen P L 2008 *Nano Lett.* **8** 2458

[59] Davidovikj D, Slim J J, Cartamil-Bueno S J, van der Zant H S J, Steeneken P G and Venstra W J 2016 *Nano Lett.* **16** 2768

[60] Ferrari P F, Kim S and van der Zande A M 2021 *Nano Lett.* **21** 8058

[61] Zhang X, Makles K, Colombier L, Metten D, Majjad H, Verlot P and Berciaud S 2020 *Nat. Commun.* **11** 5526

[62] Barton R A, Storch I R, Adiga V P, Sakakibara R, Cipriany B R, Ilic B, Wang S P, Ong P, McEuen P L, Parpia J M and Craighead H G 2012 *Nano Lett.* **12** 4681

[63] Lee S, Chen C, Deshpande V V, Lee G-H, Lee I, Lekas M, Gondarenko A, Yu Y-J, Shepard K, Kim P and Hone J 2013 *Appl. Phys. Lett.* **102** 153101

[64] Blaikie A, Miller D and Alemán B J 2019 *Nat. Commun.* **10** 4726

[65] Kumar M and Bhaskaran H 2015 *Nano Lett.* **15** 2562

[66] Reserbat-Plantey A, Marty L, Arcizet O, Bendiab N and Bouchiat V 2012 *Nat. Nanotechnol.* **7** 151

[67] Dolleman R J, Chakraborty D, Ladiges D R, van der Zant H S J, Sader J E and Steeneken P G 2021 *Nano Lett.* **21** 7617

[68] Koenig S P, Boddeti N G, Dunn M L and Bunch J S 2011 *Nat. Nanotechnol.* **6** 543

[69] Ye F, Lee J and Feng P X L 2017 *IEEE 30th International Conference on Micro Electro Mechanical Systems (MEMS)*, 22-26 Jan. 2017, p 68-71

[70] Zande A M v d, Barton R A, Alden J S, Ruiz-Vargas C S, Whitney W S, Pham P H Q, Park J, Parpia J M, Craighead H G and McEuen P L 2010 *Nano Lett.* **10** 4869

[71] Shivaraman S, Barton R A, Yu X, Alden J, Herman L, Chandrashekhar M V S, Park J, McEuen P L, Parpia J M, Craighead H G and Spencer M G 2009 *Nano Lett.* **9** 3100

[72] Castellanos-Gomez A, Buscema M, Molenaar R, Singh V, Janssen L, van der Zant H S J and Steele G A 2014 *2D Mater.* **1** 011002

[73] Suk J W, Kitt A, Magnuson C W, Hao Y, Ahmed S, An J, Swan A K, Goldberg B B and Ruoff R S 2011 *ACS Nano* **5** 6916

[74] Bao W, Miao F, Chen Z, Zhang H, Jang W, Dames C and Lau C N 2009 *Nat. Nanotechnol.* **4** 562

[75] Song X, Oksanen M, Li J, Hakonen P J and Sillanpää M A 2014 *Phys. Rev. Lett.* **113** 027404

[76] Dolleman R J, Belardinelli P, Houri S, van der Zant H S J, Alijani F and Steeneken P G 2019 *Nano Lett.* **19** 1282

[77] Suo J-J, Li W-J, Cheng Z-D, Zhao Z-F, Chen H, Li B-L, Zhou Q, Wang Y, Song H-Z, Niu X-B and Deng G-W 2022 *J. Phys.: Condens. Matter* **34** 374004

[78] Zhang Z-Z, Hu Q, Song X-X, Ying Y, Li H-O, Zhang Z and Guo G-P 2020 *Adv. Mater.* **32** 2005625

[79] Manzeli S, Dumcenco D, Migliato Marega G and Kis A 2019 *Nat. Commun.* **10** 4831

[80] Singh V, Sengupta S, Solanki H S, Dhall R, Allain A, Dhara S, Pant P and Deshmukh M M 2010

[81] Chen C 2013 *Graphene NanoElectroMechanical Resonators and Oscillators* (Dissertation/Thesis) (Columbia University)

[82] Tsioutsios I 2016 *Mechanical resonators based on graphene and carbon nanotubes* (Dissertation/Thesis) (ICFO-Institut de Ciencies Fotoniques)

[83] Samanta C, Czaplewski D A, Bonis S L D, Møller C B, Tormo-Queralt R, Miller C S, Jin Y, Pistolesi F and Bachtold A 2023 *Appl. Phys. Lett.* **123** 203502

[84] Gouttenoire V, Barois T, Perisanu S, Leclercq J-L, Purcell S T, Vincent P and Ayari A 2010 *Small* **6** 1060

[85] Zhang C, Zhang Y, Yang C, Lu H, Chen F, Yan Y and Moser J 2024 *Commun. Eng.* **3** 45

[86] Song X, Oksanen M, Sillanpää M A, Craighead H G, Parpia J M and Hakonen P J 2012 *Nano Lett.* **12** 198

[87] Singh V, Shevchuk O, Blanter Y M and Steele G A 2016 *Phys. Rev. B* **93** 245407

[88] Nishiguchi K, Yamaguchi H and Fujiwara A 2023 *Phys. Rev. Appl.* **19** L011003

[89] Han E, Yu J, Annevelink E, Son J, Kang D A, Watanabe K, Taniguchi T, Ertekin E, Huang P Y and van der Zande A M 2020 *Nat. Mater.* **19** 305

[90] Wang G, Dai Z, Xiao J, Feng S, Weng C, Liu L, Xu Z, Huang R and Zhang Z 2019 *Phys. Rev. Lett.* **123** 116101

[91] Wah T 1962 *J. Acoust. Soc. Am.* **34** 275

[92] Lee J, Wang Z, He K, Shan J and Feng P X L 2013 *ACS Nano* **7** 6086

[93] Lee J, Wang Z, He K, Yang R, Shan J and Feng P X L *Sci. Adv.* **4** eaao6653

[94] Zhu J, Xu B, Xiao F, Liang Y, Jiao C, Li J, Deng Q, Wu S, Wen T, Pei S, Xia J and Wang Z 2022 *Nano Lett.* **22** 5107

[95] Zheng X-Q, Lee J and Feng P X L 2017 *Microsyst. Nanoeng.* **3** 17038

[96] Castellanos-Gomez A, van Leeuwen R, Buscema M, van der Zant H S J, Steele G A and Venstra W J 2013 *Adv. Mater.* **25** 6719

[97] Sangani L D V, Mandal S, Ghosh S, Watanabe K, Taniguchi T and Deshmukh M M 2022 *Nano Lett.* **22** 3612

[98] Kim S, Annevelink E, Han E, Yu J, Huang P Y, Ertekin E and van der Zande A M 2020 *Nano Lett.* **20** 1201

[99] Sapmaz S, Blanter Y M, Gurevich L and van der Zant H S J 2003 *Phys. Rev. B* **67** 235414

[100] Landau L D, & Lifshitz, E. M. 1986 *Theory of Elasticity*, 3rd edn (Oxford: Elsevier)

[101] Timoshenko S 1959 *Theory of Plates and Shells*, 2nd edn (New York: McGraw-Hill)

[102] Wang Z and Feng P X-L 2014 *Appl. Phys. Lett.* **104** 103109

[103] Eichler A, Moser J, Dykman M I and Bachtold A 2013 *Nat. Commun.* **4** 2843

[104] Samanta C, Arora N and Naik A K 2018 *Appl. Phys. Lett.* **113** 113101

[105] Kozinsky I, Postma H W C, Bargatin I and Roukes M L 2006 *Appl. Phys. Lett.* **88** 253101

[106] Lifshitz R and Cross M C 2008 *Reviews of Nonlinear Dynamics and Complexity*, ed Schuster H G (Wiley) pp 1-52

[107] Cui T-H, Li J, Yuan Q, Wei Y-Q, Dai S-Q, Li P-D, Zhou F, Zhang J-Q, Chen L and Feng M 2023 *Chin. Phys. Lett.* **40** 080501

[108] Shaochun L, Tian T, Peiran Y, Pu H, Liang Z and Jiangfeng D 2021 *Chin. Phys. Lett.* **38** 020502

[109] Šiškins M, Keşkekler A, Houmes M J A, Mañas-Valero S, Koperski M, Coronado E, Blanter Y M, van der Zant H S J, Steeneken P G and Alijani F 2025 *Nat. Commun.* **16** 2177

[110]  Imboden M and Mohanty P 2014 *Phys. Rep.* **534** 89
[111]  Miao T, Yeom S, Wang P, Standley B and Bockrath M 2014 *Nano Lett.* **14** 2982
[112]  Takamura M, Okamoto H, Furukawa K, Yamaguchi H and Hibino H 2014 *J. Appl. Phys.* **116** 064304
[113]  Seoánez C, Guinea F and Castro Neto A H 2007 *Phys. Rev. B* **76** 125427
[114]  Kim S Y and Park H S 2009 *Appl. Phys. Lett.* **94** 101918
[115]  Dykman M I and Krivoglaz M A 1975 *Phys. Status Solidi B* **68** 111
[116]  Zaitsev S, Shtempluck O, Buks E and Gottlieb O 2012 *Nonlinear Dyn.* **67** 859
[117]  Croy A, Midtvedt D, Isacsson A and Kinaret J M 2012 *Phys. Rev. B* **86** 235435
[118]  De Alba R, Massel F, Storch I R, Abhilash T S, Hui A, McEuen P L, Craighead H G and Parpia J M 2016 *Nat. Nanotechnol.* **11** 741
[119]  Dolleman R J, Davidovikj D, Cartamil-Bueno S J, van der Zant H S J and Steeneken P G 2016 *Nano Lett.* **16** 568
[120]  Novotny L 2010 *Am. J. Phys.* **78** 1199
[121]  Frimmer M and Novotny L 2014 *Am. J. Phys.* **82** 947
[122]  Keşkekler A, Bos V, Aragón A M, Steeneken P G and Alijani F 2023 *Phys. Rev. Appl.* **20** 064020
[123]  Keşkekler A, Arjmandi-Tash H, Steeneken P G and Alijani F 2022 *Nano Lett.* **22** 6048
[124]  Aspelmeyer M, Kippenberg T J and Marquardt F 2014 *Rev. Mod. Phys.* **86** 1391
[125]  Dolleman R J, Houri S, Chandrashekar A, Alijani F, van der Zant H S J and Steeneken P G 2018 *Sci. Rep.* **8** 9366
[126]  Yang C, Zhang Y, Lu H, Zhang C, Chen F, Yan Y, Xue F, Eichler A and Moser J 2025 *Phys. Rev. Appl.* **23** 054036
[127]  Shi Z, Lu H, Zhang L, Yang R, Wang Y, Liu D, Guo H, Shi D, Gao H, Wang E and Zhang G 2012 *Nano Res.* **5** 82
[128]  Li P, Jing G, Zhang B, Sando S and Cui T 2014 *Appl. Phys. Lett.* **104** 113110
[129]  Muruganathan M, Van N H, Schmidt M E and Mizuta H 2022 *Adv. Funct. Mater.* **32** 2209151
[130]  Kim S M, Song E B, Lee S, Seo S, Seo D H, Hwang Y, Candler R and Wang K L 2011 *Appl. Phys. Lett.* **99** 023103
[131]  Li P, You Z, Haugstad G and Cui T 2011 *Appl. Phys. Lett.* **98** 253105
[132]  Chen M E, Rojo M M, Lian F, Koeln J, Sood A, Bohaichuk S M, Neumann C M, Garrow S G, Goodson K E, Alleyne A G and Pop E 2021 *2D Mater.* **8** 035055
[133]  Tian H, Xie D, Yang Y, Ren T-L, Wang Y-F, Zhou C-J, Peng P-G, Wang L-G and Liu L-T 2012 *Nanoscale* **4** 2272
[134]  Suk J W, Kirk K, Hao Y, Hall N A and Ruoff R S 2012 *Adv. Mater.* **24** 6342
[135]  Smith A D, Niklaus F, Paussa A, Vaziri S, Fischer A C, Sterner M, Forsberg F, Delin A, Esseni D, Palestri P, Östling M and Lemme M C 2013 *Nano Lett.* **13** 3237
[136]  Wang Q, Hong W and Dong L 2016 *Nanoscale* **8** 7663
[137]  Davidovikj D, Scheepers P H, van der Zant H S J and Steeneken P G 2017 *ACS Appl. Mater. Interfaces* **9** 43205
[138]  Šiškins M, Lee M, Wehenkel D, van Rijn R, de Jong T W, Renshof J R, Hopman B C, Peters W S J M, Davidovikj D, van der Zant H S J and Steeneken P G 2020 *Microsyst. Nanoeng.* **6** 102
[139]  Romijn J, Vollebregt S, Dolleman R J, Singh M, Zant H S J v d, Steeneken P G and Sarro P M *2018 IEEE 13th Annual International Conference on Nano/Micro Engineered and Molecular Systems (NEMS)*, 22-26 April 2018, p 11-14

[140] Zhang Z, Liu Q, Ma H, Ke N, Ding J, Zhang W and Fan X 2024 *IEEE Sens. J.* **24** 25227

[141] Lee H-L, Yang Y-C and Chang W-J 2013 *Jpn. J. Appl. Phys.* **52** 025101

[142] Ni W, Lu P, Fu X, Zhang W, Shum P P, Sun H, Yang C, Liu D and Zhang J 2018 *Opt. Express* **26** 20758

[143] Fan X, Forsberg F, Smith A D, Schröder S, Wagner S, Östling M, Lemme M C and Niklaus F 2019 *Nano Lett.* **19** 6788

[144] Ding J, Ma H, He C, Zhang W and Fan X 2025 *ACS Nano* **19** 12253

[145] Fan X, He C, Ding J, Afyouni Akbari S S and Zhang W 2024 *Microsyst. Nanoeng.* **10** 150

[146] Fan X, Moreno-Garcia D, Ding J, Gylfason K B, Villanueva L G and Niklaus F 2024 *ACS Appl. Nano Mater.* **7** 102

[147] Moreno-Garcia D, Fan X, Smith A D, Lemme M C, Messina V, Martin-Olmos C, Niklaus F and Villanueva L G 2022 *Small* **18** 2201816

[148] He C, Ding J and Fan X 2024 *Micromachines* **15** 409

[149] Liu Q, He C, Ding J, Zhang W and Fan X 2024 *ACS Appl. Mater. Interfaces* **16** 59066

[150] Chen Y-M, He S-M, Huang C-H, Huang C-C, Shih W-P, Chu C-L, Kong J, Li J and Su C-Y 2016 *Nanoscale* **8** 3555

[151] Berger C, Phillips R, Centeno A, Zurutuza A and Vijayaraghavan A 2017 *Nanoscale* **9** 17439

[152] Smith A D, Niklaus F, Paussa A, Schröder S, Fischer A C, Sterner M, Wagner S, Vaziri S, Forsberg F, Esseni D, Östling M and Lemme M C 2016 *ACS Nano* **10** 9879

[153] Smith A D, Elgammal K, Fan X, Lemme M C, Delin A, Råsander M, Bergqvist L, Schröder S, Fischer A C, Niklaus F and Östling M 2017 *RSC Adv.* **7** 22329

[154] Zhang Z, Liu Q, Gao Q, Si F, Cao H, Ding J, Zhang W and Fan X 2025 *IEEE Sens. J.* (Early Access)

[155] Rosłoń I E, Japaridze A, Steeneken P G, Dekker C and Alijani F 2022 *Nat. Nanotechnol.* **17** 637

[156] Maldovan M 2013 *Nature* **503** 209

[157] Hatanaka D, Mahboob I, Onomitsu K and Yamaguchi H 2014 *Nat. Nanotechnol.* **9** 520

[158] Cha J and Daraio C 2018 *Nat. Nanotechnol.* **13** 1016

[159] Kurosu M, Hatanaka D, Onomitsu K and Yamaguchi H 2018 *Nat. Commun.* **9** 1331

[160] Ghadimi A H, Fedorov S A, Engelsen N J, Bereyhi M J, Schilling R, Wilson D J and Kippenberg T J 2018 *Science* **360** 764

[161] MacCabe G S, Ren H, Luo J, Cohen J D, Zhou H, Sipahigil A, Mirhosseini M and Painter O 2020 *Science* **370** 840

[162] Tsaturyan Y, Barg A, Polzik E S and Schliesser A 2017 *Nat. Nanotechnol.* **12** 776

[163] Kalaee M, Mirhosseini M, Dieterle P B, Peruzzo M, Fink J M and Painter O 2019 *Nat. Nanotechnol.* **14** 334

[164] Beccari A, Visani D A, Fedorov S A, Bereyhi M J, Boureau V, Engelsen N J and Kippenberg T J 2022 *Nat. Phys.* **18** 436

[165] Engelsen N J, Beccari A and Kippenberg T J 2024 *Nat. Nanotechnol.* **19** 725

[166] Serra-Garcia M, Peri V, Süsstrunk R, Bilal O R, Larsen T, Villanueva L G and Huber S D 2018 *Nature* **555** 342

[167] He H, Qiu C, Ye L, Cai X, Fan X, Ke M, Zhang F and Liu Z 2018 *Nature* **560** 61

[168] Cha J, Kim K W and Daraio C 2018 *Nature* **564** 229

[169] Zhang X, Xiao M, Cheng Y, Lu M-H and Christensen J 2018 *Commun. Phys.* **1** 97

[170] Ma G, Xiao M and Chan C T 2019 *Nat. Rev. Phys.* **1** 281

[171] Yu Y, Kirchhof J N, Tsarapkin A, Deinhart V, Yücel O, Höfer B, Höflich K and Bolotin K I 2023 *2D*

*Mater.* **10** 045012

[172] Kirchhof J N and Bolotin K I 2023 *npj 2D Mater. Appl.* **7** 10

[173] Zhang Z-D, Lu M-H and Chen Y-F 2024 *Phys. Rev. Lett.* **132** 086302

[174] Rugar D, Budakian R, Mamin H J and Chui B W 2004 *Nature* **430** 329

[175] Reserbat-Plantey A, Schädler K G, Gaudreau L, Navickaite G, Güttinger J, Chang D, Toninelli C, Bachtold A and Koppens F H L 2016 *Nat. Commun.* **7** 10218

[176] Bolotin K I, Ghahari F, Shulman M D, Stormer H L and Kim P 2009 *Nature* **462** 196

[177] Du X, Skachko I, Duerr F, Luican A and Andrei E Y 2009 *Nature* **462** 192

[178] Tsui Y-C, He M, Hu Y, Lake E, Wang T, Watanabe K, Taniguchi T, Zaletel M P and Yazdani A 2024 *Nature* **628** 287

[179] Khivrich I, Clerk A A and Ilani S 2019 *Nat. Nanotechnol.* **14** 161

[180] Urgell C, Yang W, De Bonis S L, Samanta C, Esplandiu M J, Dong Q, Jin Y and Bachtold A 2020 *Nat. Phys.* **16** 32

[181] Wen Y, Ares N, Schupp F J, Pei T, Briggs G A D and Laird E A 2020 *Nat. Phys.* **16** 75

[182] Samanta C, De Bonis S L, Møller C B, Tormo-Queralt R, Yang W, Urgell C, Stamenic B, Thibeault B, Jin Y, Czaplewski D A, Pistolesi F and Bachtold A 2023 *Nat. Phys.* **19** 1340

[183] Wang X, Cong L, Zhu D, Yuan Z, Lin X, Zhao W, Bai Z, Liang W, Sun X, Deng G-W and Jiang K 2021 *Nano Res.* **14** 1156

[184] Frisenda R, Navarro-Moratalla E, Gant P, Pérez De Lara D, Jarillo-Herrero P, Gorbachev R V and Castellanos-Gomez A 2018 *Chem. Soc. Rev* **47** 53

[185] Schneider G F, Calado V E, Zandbergen H, Vandersypen L M K and Dekker C 2010 *Nano Lett.* **10** 1912

[186] Yang X, Li J, Song R, Zhao B, Tang J, Kong L, Huang H, Zhang Z, Liao L, Liu Y, Duan X and Duan X 2023 *Nat. Nanotechnol.* **18** 471

[187] Wang L, Meric I, Huang P Y, Gao Q, Gao Y, Tran H, Taniguchi T, Watanabe K, Campos L M, Muller D A, Guo J, Kim P, Hone J, Shepard K L and Dean C R 2013 *Science* **342** 614

[188] Pizzocchero F, Gammelgaard L, Jessen B S, Caridad J M, Wang L, Hone J, Bøggild P and Booth T J 2016 *Nat. Commun.* **7** 11894

[189] Dean C R, Young A F, Meric I, Lee C, Wang L, Sorgenfrei S, Watanabe K, Taniguchi T, Kim P, Shepard K L and Hone J 2010 *Nat. Nanotechnol.* **5** 722

[190] Zomer P J, Dash S P, Tombros N and van Wees B J 2011 *Appl. Phys. Lett.* **99** 232104

[191] Purdie D G, Pugno N M, Taniguchi T, Watanabe K, Ferrari A C and Lombardo A 2018 *Nat. Commun.* **9** 5387

[192] Zhao Y, Song Y, Hu Z, Wang W, Chang Z, Zhang Y, Lu Q, Wu H, Liao J, Zou W, Gao X, Jia K, Zhuo L, Hu J, Xie Q, Zhang R, Wang X, Sun L, Li F, Zheng L, Wang M, Yang J, Mao B, Fang T, Wang F, Zhong H, Liu W, Yan R, Yin J, Zhang Y, Wei Y, Peng H, Lin L and Liu Z 2022 *Nat. Commun.* **13** 4409

[193] Wang W, Clark N, Hamer M, Carl A, Tovari E, Sullivan-Allsop S, Tillotson E, Gao Y, de Latour H, Selles F, Howarth J, Castanon E G, Zhou M, Bai H, Li X, Weston A, Watanabe K, Taniguchi T, Mattevi C, Bointon T H, Wiper P V, Strudwick A J, Ponomarenko L A, Kretinin A V, Haigh S J, Summerfield A and Gorbachev R 2023 *Nat. Electron.* **6** 981

[194] Jaikissoon M, Köroğlu Ç, Yang J A, Neilson K, Saraswat K C and Pop E 2024 *Nat. Electron.* **7** 885

[195] Lu H, Yang C, Tian Y, Lu J, Xu F, Zhang C, Chen F, Ying Y, Schädler K G, Wang C, Koppens F H L, Reserbat-Plantey A and Moser J 2021 *Precis. Eng.* **72** 769

[196] Benameur M M, Gargiulo F, Manzeli S, Autès G, Tosun M, Yazyev O V and Kis A 2015 *Nat. Commun.* **6** 8582

[197] Nicholl R J T, Conley H J, Lavrik N V, Vlassiouk I, Puzyrev Y S, Sreenivas V P, Pantelides S T and Bolotin K I 2015 *Nat. Commun.* **6** 8789

[198] Dai Z, Liu L and Zhang Z 2019 *Adv. Mater.* **31** 1805417

[199] Hou Y, Zhou J, He Z, Chen J, Zhu M, Wu H and Lu Y 2024 *Nat. Commun.* **15** 4033

[200] Deng S and Berry V 2016 *Mater. Today* **19** 197

[201] Lu H, Yang C, Zhang C, Zhang Y, Chen F, Ying Y, Zhang Z-Z, Song X-X, Deng G-W, Yan Y and Moser J 2025 *arXiv e-prints* arXiv:2502.19783

[202] Alden J S, Tsen A W, Huang P Y, Hovden R, Brown L, Park J, Muller D A and McEuen P L 2013 *Proc. Natl. Acad. Sci.* **110** 11256

[203] McGilly L J, Kerelsky A, Finney N R, Shapovalov K, Shih E-M, Ghiotto A, Zeng Y, Moore S L, Wu W, Bai Y, Watanabe K, Taniguchi T, Stengel M, Zhou L, Hone J, Zhu X, Basov D N, Dean C, Dreyer C E and Pasupathy A N 2020 *Nat. Nanotechnol.* **15** 580

[204] Kazmierczak N P, Van Winkle M, Ophus C, Bustillo K C, Carr S, Brown H G, Ciston J, Taniguchi T, Watanabe K and Bediako D K 2021 *Nat. Mater.* **20** 956

[205] Lau C N, Bockrath M W, Mak K F and Zhang F 2022 *Nature* **602** 41

[206] Amontree J, Yan X, DiMarco C S, Levesque P L, Adel T, Pack J, Holbrook M, Cupo C, Wang Z, Sun D, Biacchi A J, Wilson-Stokes C E, Watanabe K, Taniguchi T, Dean C R, Hight Walker A R, Barmak K, Martel R and Hone J 2024 *Nature* **630** 636

[207] Li X, Cai W, An J, Kim S, Nah J, Yang D, Piner R, Velamakanni A, Jung I, Tutuc E, Banerjee S K, Colombo L and Ruoff R S 2009 *Science* **324** 1312

[208] Lee J-H, Lee E K, Joo W-J, Jang Y, Kim B-S, Lim J Y, Choi S-H, Ahn S J, Ahn J R, Park M-H, Yang C-W, Choi B L, Hwang S-W and Whang D 2014 *Science* **344** 286

[209] Wang M, Huang M, Luo D, Li Y, Choe M, Seong W K, Kim M, Jin S, Wang M, Chatterjee S, Kwon Y, Lee Z and Ruoff R S 2021 *Nature* **596** 519

[210] Novoselov K S, Mishchenko A, Carvalho A and Castro Neto A H 2016 *Science* **353** aac9439

[211] Will M, Hamer M, Müller M, Noury A, Weber P, Bachtold A, Gorbachev R V, Stampfer C and Güttinger J 2017 *Nano Lett.* **17** 5950

[212] Singh R, Nicholl R J T, Bolotin K I and Ghosh S 2018 *Nano Lett.* **18** 6719

[213] Kim S, Yu J and van der Zande A M 2018 *Nano Lett.* **18** 6686

[214] Lee S, Adiga V P, Barton R A, van der Zande A M, Lee G-H, Ilic B R, Gondarenko A, Parpia J M, Craighead H G and Hone J 2013 *Nano Lett.* **13** 4275

[215] Kumar R, Session D W, Tsuchikawa R, Homer M, Paas H, Watanabe K, Taniguchi T and Deshpande V V 2020 *Appl. Phys. Lett.* **117** 183103

[216] Adiga V P, De Alba R, Storch I R, Yu P A, Ilic B, Barton R A, Lee S, Hone J, McEuen P L, Parpia J M and Craighead H G 2013 *Appl. Phys. Lett.* **103** 143103

[217] Jiang S, Xie H, Shan J and Mak K F 2020 *Nat. Mater.* **19** 1295

[218] Singh R, Sarkar A, Guria C, Nicholl R J T, Chakraborty S, Bolotin K I and Ghosh S 2020 *Nano Lett.* **20** 4659

[219] Ye F, Islam A, Zhang T and Feng P X L 2021 *Nano Lett.* **21** 5508

[220] Miller D, Blaikie A and Alemán B J 2020 *Nano Lett.* **20** 2378

[221] Goldsche M, Sonntag J, Khodkov T, Verbiest G J, Reichardt S, Neumann C, Ouaj T, von den Driesch N, Buca D and Stampfer C 2018 *Nano Lett.* **18** 1707

[222] Schwarz C, Pigeau B, Mercier de Lépinay L, Kuhn A G, Kalita D, Bendiab N, Marty L, Bouchiat V and Arcizet O 2016 *Phys. Rev. Appl.* **6** 064021

[223] Zhang T, Wang H, Xia X, Yan N, Sha X, Huang J, Watanabe K, Taniguchi T, Zhu M, Wang L, Gao J, Liang X, Qin C, Xiao L, Sun D, Zhang J, Han Z and Li X 2022 *Light Sci. Appl.* **11** 48

[224] Zeng Q-Y, Su G-X, Song A-S, Mei X-Y, Xu Z-Y, Ying Y, Zhang Z-Z, Song X-X, Deng G-W, Moser J, Ma T-B, Tan P-H and Zhang X 2025 *Nat. Commun.* **16** 3793

[225] Wopereis M P F, Bouman N, Dutta S, Steeneken P G, Alijani F and Verbiest G J 2024 *J. Appl. Phys.* **136** 014302

[226] Verbiest G J, Goldsche M, Sonntag J, Khodkov T, von den Driesch N, Buca D and Stampfer C 2021 *2D Mater.* **8** 035039

[227] Pistolesi F, Cleland A N and Bachtold A 2021 *Phys. Rev. X* **11** 031027

[228] Yang Y, Kladarić I, Drimmer M, von Lüpke U, Lenterman D, Bus J, Marti S, Fadel M and Chu Y 2024 *Science* **386** 783

[229] Lee J, LaHaye M D and Feng P X-L 2022 *Appl. Phys. Lett.* **120** 014001

[230] Allen M T, Martin J and Yacoby A 2012 *Nat. Commun.* **3** 934

[231] Song X-X, Li H-O, You J, Han T-Y, Cao G, Tu T, Xiao M, Guo G-C, Jiang H-W and Guo G-P 2015 *Sci. Rep.* **5** 8142

[232] Jing F-M, Zhang Z-Z, Qin G-Q, Luo G, Cao G, Li H-O, Song X-X and Guo G-P 2022 *Adv. Quantum Technol.* **5** 2100162

[233] Eich M, Herman F, Pisoni R, Overweg H, Kurzmann A, Lee Y, Rickhaus P, Watanabe K, Taniguchi T, Sigrist M, Ihn T and Ensslin K 2018 *Phys. Rev. X* **8** 031023

[234] Banszerus L, Möller S, Hecker K, Icking E, Watanabe K, Taniguchi T, Hassler F, Volk C and Stampfer C 2023 *Nature* **618** 51

[235] Banszerus L, Rothstein A, Fabian T, Möller S, Icking E, Trellenkamp S, Lentz F, Neumaier D, Watanabe K, Taniguchi T, Libisch F, Volk C and Stampfer C 2020 *Nano Lett.* **20** 7709

[236] Tong C, Garreis R, Knothe A, Eich M, Sacchi A, Watanabe K, Taniguchi T, Fal'ko V, Ihn T, Ensslin K and Kurzmann A 2021 *Nano Lett.* **21** 1068

[237] Banszerus L, Rothstein A, Icking E, Möller S, Watanabe K, Taniguchi T, Stampfer C and Volk C 2021 *Appl. Phys. Lett.* **118** 103101

[238] Jing F-M, Qin G-Q, Zhang Z-Z, Song X-X and Guo G-P 2023 *Appl. Phys. Lett.* **123** 184001

[239] Banszerus L, Möller S, Steiner C, Icking E, Trellenkamp S, Lentz F, Watanabe K, Taniguchi T, Volk C and Stampfer C 2021 *Nat. Commun.* **12** 5250

[240] Kurzmann A, Kleeorin Y, Tong C, Garreis R, Knothe A, Eich M, Mittag C, Gold C, de Vries F K, Watanabe K, Taniguchi T, Fal'ko V, Meir Y, Ihn T and Ensslin K 2021 *Nat. Commun.* **12** 6004

[241] Duprez H, Cances S, Omahen A, Masseroni M, Ruckriegel M J, Adam C, Tong C, Garreis R, Gerber J D, Huang W, Gächter L, Watanabe K, Taniguchi T, Ihn T and Ensslin K 2024 *Nat. Commun.* **15** 9717

[242] Möller S, Banszerus L, Knothe A, Steiner C, Icking E, Trellenkamp S, Lentz F, Watanabe K, Taniguchi T, Glazman L I, Fal'ko V I, Volk C and Stampfer C 2021 *Phys. Rev. Lett.* **127** 256802

[243] Jing F-M, Shen Z-X, Qin G-Q, Zhang W-K, Lin T, Cai R, Zhang Z-Z, Cao G, He L, Song X-X and Guo G-P 2025 *Phys. Rev. Appl.* **23** 044053

[244] Hecker K, Banszerus L, Schäpers A, Möller S, Peters A, Icking E, Watanabe K, Taniguchi T, Volk C and Stampfer C 2023 *Nat. Commun.* **14** 7911

[245] Banszerus L, Hecker K, Möller S, Icking E, Watanabe K, Taniguchi T, Volk C and Stampfer C

2022 *Nat. Commun.* **13** 3637

[246] Gächter L M, Garreis R, Gerber J D, Ruckriegel M J, Tong C, Kratochwil B, de Vries F K, Kurzmann A, Watanabe K, Taniguchi T, Ihn T, Ensslin K and Huang W W 2022 *PRX Quantum* **3** 020343

[247] Sierra J F, Fabian J, Kawakami R K, Roche S and Valenzuela S O 2021 *Nat. Nanotechnol.* **16** 856

[248] Yang H, Valenzuela S O, Chshiev M, Couet S, Dieny B, Dlubak B, Fert A, Garello K, Jamet M, Jeong D-E, Lee K, Lee T, Martin M-B, Kar G S, Sénéor P, Shin H-J and Roche S 2022 *Nature* **606** 663

[249] Bisswanger T, Winter Z, Schmidt A, Volmer F, Watanabe K, Taniguchi T, Stampfer C and Beschoten B 2022 *Nano Lett.* **22** 4949

[250] Luo Z, Yu Z, Lu X, Niu W, Yu Y, Yao Y, Tian F, Tan C L, Sun H, Gao L, Qin W, Xu Y, Zhao Q and Song X-X 2024 *Nano Lett.* **24** 6183

[251] Garreis R, Tong C, Terle J, Ruckriegel M J, Gerber J D, Gächter L M, Watanabe K, Taniguchi T, Ihn T, Ensslin K and Huang W W 2024 *Nat. Phys.* **20** 428

[252] Tong C, Kurzmann A, Garreis R, Huang W W, Jele S, Eich M, Ginzburg L, Mittag C, Watanabe K, Taniguchi T, Ensslin K and Ihn T 2022 *Phys. Rev. Lett.* **128** 067702

[253] Yin J, Tan C, Barcons-Ruiz D, Torre I, Watanabe K, Taniguchi T, Song J C W, Hone J and Koppens F H L 2022 *Science* **375** 1398

[254] Li J, Zhang R-X, Yin Z, Zhang J, Watanabe K, Taniguchi T, Liu C and Zhu J 2018 *Science* **362** 1149

[255] Garreis R, Knothe A, Tong C, Eich M, Gold C, Watanabe K, Taniguchi T, Fal'ko V, Ihn T, Ensslin K and Kurzmann A 2021 *Phys. Rev. Lett.* **126** 147703

[256] Qin G-Q, Jing F-M, Hao T-Y, Jiang S-L, Zhang Z-Z, Cao G, Song X-X and Guo G-P 2025 *Phys. Rev. Lett.* **134** 036301

[257] Li H-K, Fong K Y, Zhu H, Li Q, Wang S, Yang S, Wang Y and Zhang X 2019 *Nat. Photonics* **13** 397

[258] Fedele F, Cerisola F, Bresque L, Vigneau F, Monsel J, Tabanera J, Aggarwal K, Dexter J, Sevitz S, Dunlop J, Auffèves A, Parrondo J, Pályi A, Anders J and Ares N 2024 *arXiv e-prints* arXiv:2402.19288